\documentclass[pra,twocolumn,pra,groupedaddress]{revtex4-1}

\usepackage{graphicx}% Include figure files
\usepackage{dcolumn}% Align table columns on decimal point
\usepackage{bm}% bold math
\usepackage{epsfig}
\usepackage{amsmath}
\usepackage{amssymb}
\usepackage{color}
\usepackage{bbm}
\usepackage{braket}
\usepackage{float}

\usepackage[colorlinks=true,citecolor=blue,linkcolor=blue,urlcolor=blue]{hyperref}

\begin{document}

\title{Chiral quantum optics in the bulk of photonic quantum Hall systems}

\author{Daniele De Bernardis$^1$, Francesco S. Piccioli$^1$, Peter Rabl$^{2,3,4,5}$, and Iacopo Carusotto$^1$}
\affiliation{$^1$Pitaevskii BEC Center, CNR-INO and Dipartimento di Fisica, Università di Trento, I-38123 Trento, Italy}
\affiliation{$^2$Vienna Center for Quantum Science and Technology, Atominstitut, TU Wien, 1020 Vienna, Austria}
\affiliation{$^3$Technical University of Munich, TUM School of Natural Sciences, Physics Department, 85748 Garching, Germany} 
\affiliation{$^4$Walther-Meißner-Institut, Bayerische Akademie der Wissenschaften, 85748 Garching, Germany}
\affiliation{$^5$Munich Center for Quantum Science and Technology (MCQST), 80799 Munich, Germany}

\date{\today}

%%%%
\begin{abstract} 
We study light-matter interactions in the bulk of a two-dimensional photonic lattice system, where photons are subject to the combined effect of a synthetic magnetic field and an orthogonal synthetic electric field. In this configuration, chiral waveguide modes appear in the bulk region of the lattice, in direct analogy to transverse Hall currents in electronic systems. By evaluating the non-Markovian dynamics of emitters that are coupled to those modes, we identify critical coupling conditions, under which the shape of the spontaneously emitted photons becomes almost fully symmetric. Combined with a directional,  dispersionless propagation, this property enables a complete reabsorption of the photon by another distant emitter, without relying on any time-dependent control.  We show that this mechanism can be generalized to arbitrary in-plane synthetic potentials, thereby enabling flexible realizations of re-configurable networks of quantum emitters with arbitrary chiral connectivity.
\end{abstract}
 
\maketitle

\section{Introduction}

The challenge of building fully operative quantum devices such as quantum computers, quantum simulators and quantum cryptography systems has stimulated an unprecedented flow of ideas for the implementation of technologies based on the principles of quantum mechanics \cite{zoller_quantum_world_2_doi:10.1142/p983}.
In this effort,  many disruptive ideas came by combining concepts from different areas of quantum sciences \cite{Rabl_PNAS_pnas.1419326112}. Theories and concepts that were originally developed to explain fundamental physical phenomena are now re-elaborated in a new technological perspective. This process not only serves to inspire the realization of new devices, but also provides new insights on existing knowledge and contributes to building a more complete understanding of the microscopic world.

This is the case, for instance, for the quantum Hall effect. This effect was first discovered in electronic materials more than $40$ years ago \cite{Klitzing_PhysRevLett.45.494} and sparkled the development of the theory of topological materials \cite{Niu_PhysRevB.31.3372} and the proposal of novel schemes for topologically protected quantum computing~\cite{Nayak:RMP2008}. 
These concepts are now being re-elaborated in the context of photonic systems, giving rise to the field of \emph{topological photonics} \cite{ozawa_RevModPhys.91.015006}.
Here, ideas and phenomena of the integer and fractional quantum Hall effect are implemented and generalised to various synthetic photonic, phononic, atomic or even molecular platforms \cite{mohammad_edge_modes_silicon, Bloch_PhysRevLett.111.185301, Salerno_2015, Peano_PhysRevX.5.031011, Peter_PhysRevA.91.053617, Mancini_doi:10.1126/science.aaa8736, Stuhl_doi:10.1126/science.aaa8515, Simon_nature2016, OzawaPrice_NatRev_2019, Carusotto_NatRev_2020, Weber:2022ihh}, with unprecedented freedom in tuning the physical parameters and measuring observables, which were previously inaccessible in traditional solid-state systems.

These developments in fundamental science have naturally opened the way to technological applications, for instance to exploit the topologically protected chiral edge modes to create new integrated solutions for a unidirectional transport of information, robust against system imperfections~\cite{Wang_2009_Nature_topoEdgeNoBackScatt, Demler_Hafezi_Nature_2011}. 
The potential of these new devices became particularly evident in the framework of the field of \emph{chiral quantum optics} \cite{Lodahl_chiralqauntumOptics}: here, the use of topological chiral channels for the propagation of photons combined with non-linear quantum emitters, such as atoms or quantum dots, opens the way toward the creation of a full cascaded quantum network \cite{Yao_topospinedge_2013, Dlaska_2017}, which is a central piece in the development of quantum information technologies \cite{Cirac_PhysRevLett.78.3221}.

This sparkled  a new era for topological photonics experiments, with the objective of creating hybrid qubit-photonic lattice platforms in different spectral regions, from the GHz up to the optical range.
In these systems, topological features are exploited to realize complete chiral quantum optical setups, coupling their topological chiral edge channels to localized quantum emitters (or qubits) \cite{Hafezi_2018_doi:10.1126/science.aaq0327, Simon_chiralQuantumOptics_2022, Hallett_ACS2022}. In parallel to such intense experimental efforts, new innovative theoretical proposals are constantly made to exploit these devices for new technological applications \cite{Stannigel_2012, Ringel_2014, Pichler_PhysRevA.91.042116, Lemonde_2019, bello_doi:10.1126/sciadv.aaw0297, Longhi_PhysRevA.100.022123, Calajo_Mhmoodian_PhysRevX.10.031011, Longhi_photonics7010018, Yelin_PhysRevLett.124.083603, Alejandro_https://doi.org/10.48550/arxiv.2207.02090, Jiang_10.3389/fphy.2022.845579, Keyu_PhysRevLett.128.203602, Shanhui_PhysRevB.103.125423, Rabl_PhysRevLett.130.050801}.

In this article we study the light-matter interaction dynamics of two-level quantum emitters coupled to a 2D photonic lattice subject to an homogeneous perpendicular synthetic magnetic field and an in-plane homogeneous synthetic electric field. Differently from the existing chiral quantum optics literature~\cite{Yao_topospinedge_2013, Dlaska_2017, Lemonde_2019, Alejandro_https://doi.org/10.48550/arxiv.2207.02090, Longhi_photonics7010018, Longhi_PhysRevA.100.022123, mohammad_edge_modes_silicon}, which mostly focuses on light propagating along edge modes, here we investigate new strategies based on light propagation through the {\em bulk} of a 2D photonic system via the photonic analog of the Hall current. In the last decade, related anomalous transport and Berry curvature effects in the bulk of photonic systems have been the subject of several theoretical~\cite{Ozawa_PhysRevLett.112.133902,Ozawa_PhysRevB.97.201117,Brosco_PhysRevA.103.063518} and experimental~\cite{Wimmer_nat_phys_2017,Gianfrate2020_nature, Amo_PhysRevLett.126.127403} works, but have never been proposed as the operating principle of photonic devices.     

Specifically, we show here how the combined effect of crossed synthetic electric and magnetic field produces effective 1D waveguides based on the Hall effect, which allow light to unidirectionally propagate through the bulk of the lattice, similar to Hall currents. The highly in-homogeneous local density of photonic states of these waveguides makes the emission dynamics of two-level quantum emitters strongly non-Markovian. 

This intrinsic non-Markovianity is a unique feature of this system and completely modifies the nature of the most basic light-matter interaction processes such as spontaneous emission and absorption. 
The photon generated by this exotic non-Markovian spontaneous emission process naturally propagates along a single direction with a highly symmetric wavepacket shape, very different from the usual highly asymmetric wavepacket of  standard Markovian emission. 
As a direct consequence of this symmetric shape, an efficient chiral state transfer between distant emitters located in the bulk of the lattice is possible without the need for fine-tuned time-dependent control pulses~\cite{Cirac_PhysRevLett.78.3221, ZOLLER_PhysRevA.84.042341}.
As another surprising consequence of the non-Markovian light-matter interactions, this system supports atom-photon bound states in the chiral continuum \cite{BIC_review, Calajo_PhysRevLett.122.073601}, which are typically absent in the usual chiral quantum optics configurations.

In view of the high tunability of topological photonic structures,  
{these new physical effects  naturally call for technological applications and pave the way towards cascaded quantum networks with an improved performance compared to conventional setups. 
%In particular,
On top of this, our proposed approach makes full use of the whole bulk of the photonic lattice and thereby drastically increases the number of emitters that can be interfaced in a directional manner.  Thanks to the topological origin of the underlying mechanism, our proposed transfer method is also
highly robust with respect to imperfections and can be readily generalized to non-uniform electric field profiles leading to arbitrary curvilinear chiral 1D waveguides. This is of utmost interest in view of realizing reconfigurable cascaded networks with arbitrary connectivity.}

{
%Aside of these application perspectives, our results have also large fundamental interest. The intrinsic non-Markovian dynamics arising from the interplay between light-matter interactions and the photonic quantum Hall effect is a unique feature of this system, leading to completely new rules for the spontaneous emission and absorption, which are the basics building blocks of light-matter interactions. 
%Apart for the previously mentioned technical impact, this change of paradigms also lead to new exotic phenomena such as the presence of atom-photon bound states in the chiral continuum \cite{BIC_review, Calajo_PhysRevLett.122.073601}, which are typically not present in configurations when photons propagate chirally.
  
%Finally, our approach suggests a novel way to generalize and study the chiral edge dynamics in the bulk of a fully photonic platform opening up to the intriguing ideas related to 
From a fundamental science perspective, such a network could also be seen as a quantum simulator of 
the percolation theory of the quantum Hall effect and its random network representation \cite{Chalker_1988,KRAMER2005211}, a point of view which is closely connected to the properties of quantum Hall extended states \cite{Trugman_PhysRevB.27.7539,Arovas_PhysRevLett.60.619,Huo_extstates_PhysRevLett.68.1375}.
Through our findings, this physics can be now simulated with a full freedom in the choice of parameters and geometry. 
In addition, introducing non-linear emitters will provide effective interaction between photons, which may open to studies of the propagation dynamics of fractional quantum Hall edge excitations in a novel context.}

The article is structured as follows. In Sec.~\ref{sec:model} we introduce the model for our 2D photonic quantum Hall system with synthetic magnetic and synthetic electric fields. In Sec.~\ref{sec:photonic_transport_quantumHall} we use the lattice photonic Green's function to provide a basic description of photon propagation across the bulk of this system and for the appearance of effective chiral waveguide modes.
In Sec. \ref{sec:3_regimes_LM_int} we study the dynamics of a single emitter coupled to the photonic lattice and discuss the different regimes of light-matter interactions in this setup. In Sec. \ref{sec:periodic_rev_ballistic} we show how the non-Markovian emission dynamics in the critical-coupling regime enables high-fidelity quantum-state transfer operations between two emitters in the lattice. In Sec. \ref{sec:percolation_network} we extend our results to arbitrary electric field profiles and establish the concept of photonic percolation quantum networks.
Finally, in Sec. \ref{sec:conclusions} we summarize our main conclusions.

\section{The Model}
\label{sec:model}

\begin{figure}
    \centering
    \includegraphics[width=\columnwidth]{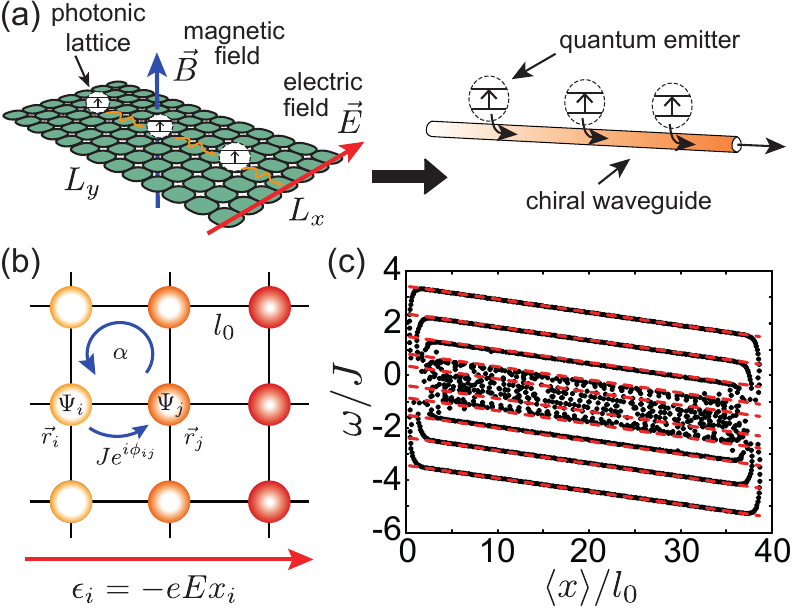}
    \caption{Photonic quantum Hall system. (a) Sketch of a 2D photonic lattice with a perpendicular synthetic magnetic field and an in-plane synthetic electric field. In analogy to Hall currents, this configuration results in the formation of chiral waveguide modes. Coupling of two-level emitters to the chiral waveguide modes leads to the directional emission and reabsorption of photons. (b) Schematic view of the photonic hopping amplitudes for the specific Landau-gauge, Harper-Hofstadter lattice configuration under consideration. (c) Spectrum of the photonic lattice Hamiltonian $H_{\rm ph}$ in Eq.~\eqref{eq:Ham_photonic_general}, where the energy of each mode is plotted as a function of the mean displacement of its center of mass along the $x$-direction. In this representation, the tilting of the Landau levels by the electric field and the presence of edge modes at the boundaries are clearly visible. The parameters used for this plot are $\alpha = 1/10$, $N_x = N_y = 40$ and $U_0/J = 0.05$ and we have assumed periodic boundary conditions (PBC) along the $y$-direction.}
    \label{fig:1}
\end{figure}
\subsection{Light-matter interactions in photonic lattices}
We consider the system depicted in Fig.~\ref{fig:1} (a), where $N$ (artificial) two-level emitters with frequency $\omega_e$ are locally coupled to a two-dimensional photonic lattice with dimensions $L_x$ and $L_y$. We denote the position of the $i$-th lattice site by $\vec{r}_i = (x_i, y_i)$ and consider a simple square lattice geometry with lattice spacing $l_0$ and $M=L_x L_y/l_0^2$ lattice sites in total. We also assume that the number of emitters is much smaller than the number of lattice sites, $N \ll M$.  

As shown in Fig.~\ref{fig:1} (b), every single lattice site represents a localized photonic mode with frequency $\epsilon_i$ and annihilation operator $\Psi_i \equiv \Psi (\vec{r}_i)$. By considering only nearest-neighbor hopping between the localized modes, the general lattice Hamiltonian is given by
 \begin{equation}\label{eq:Ham_photonic_general}
     H_{\rm ph} = \sum_{i=1}^M \epsilon_i \Psi_i^{\dag} \Psi_i - \hbar J\sum_{\braket{ij}} \left( e^{i \phi_{ij}}\Psi_i^{\dag} \Psi_j + {\rm H.c.} \right).
 \end{equation}
 The tunneling amplitude is complex-valued, with a non-trivial phase $\phi_{ij}$ to break time-reversal symmetry. 
 In our system we use this phase to generate a homogeneous synthetic magnetic field for photons by imposing $\phi_{ij} = \frac{e}{\hbar} \int_{\vec{r}_j}^{\vec{r}_i} \vec{A}(\vec{r})\cdot d\vec{r}$ \cite{ozawa_RevModPhys.91.015006}, with $\vec{A}(\vec{r}) = B(0, x,0)$ being the synthetic vector potential for a homogeneous synthetic magnetic field taken for simplicity in the Landau gauge.
As usual, we express the strength of the magnetic field in terms of the dimensionless parameter  $\alpha=e\Phi/(2\pi\hbar)$, where $\Phi= B l_0^2$ is the flux enclosed in a single plaquette.
Going beyond the setup considered in Ref. \cite{DeBernardis_PhysRevLett.126.103603}, here we impose an additional linear frequency gradient, 
\begin{equation}
    \epsilon_i = - eE\,x_i + \hbar \omega_p,
\end{equation}
where $\omega_p$ is the bare frequency of the local modes.
This simulates the effect of a homogeneous electric field in the $x$-direction. The strength of field is characterized by a voltage drop $U_{0} = eE l_0$ between two neighboring sites.  
This situation is then similar to a solid-state system, where the effect of a crossed magnetic field $\vec{B}$ and an electric field $\vec{E}$ gives rise to a quantized Hall current $\vec{\mathcal{J}}_{\rm H} \sim \vec{E} \times \vec{B}$, which flows, in our convention, along the $y$-axis. 

 Including the emitters and their coupling to the local photon modes, the total Hamiltonian for this setup is given by 
 \begin{equation}\label{eq:ham_total_light-matter}
     H = H_{\rm ph} + \sum_{n=1}^N \frac{\hbar \omega_e^n}{2}\sigma_z^n + \sum_{n=1}^N \left( \frac{\hbar g_n}{2}\Psi (\vec{r}_e^{\,n}) \sigma_+^n + {\rm H.c.} \right).
 \end{equation}
 Here, the $\sigma_{z/\pm}^n$ are Pauli matrices for an emitter at position $\vec{r}_e^{\,n}$, while $\omega_e^n\approx \omega_p$ and $g_n$ are its transition frequency and the strength of the light-matter coupling, respectively.

\subsection{A photonic lattice in the quantum Hall regime}

Since the photons are noninteracting, the properties of the photonic lattice are fully encoded in the eigenfrequencies $\omega_\lambda$ and eigenmodes $f_{\lambda}(\vec{r}_i )$ of the hopping matrix, i.e., in the solutions of the eigenvalue equation
 \begin{equation}\label{eq:photonic_eigenvalue_eq}
     \sum_j \left( \epsilon_i \delta_{ij} - \hbar J e^{i \phi_{ij}} \delta_{\braket{ij}} \right)f_{\lambda}(\vec{r}_j ) = \hbar \omega_{\lambda} f_{\lambda}(\vec{r}_i ),
 \end{equation}
 where $\delta_{ij}$ is the Kronecker delta, and $\delta_{\braket{ij}}$ is the Kronecker delta for nearest neighbors.

For a non-zero magnetic flux, $\alpha\neq 0$, and a homogeneous on-site frequency $\epsilon_i = \hbar \omega_p$ for all lattice sites, the photonic spectrum is given by the famous Hofstadter butterfly \cite{Hofstadter_PhysRevB.14.2239}, where all the eigenvalues are grouped in a finite number of narrow Landau levels, with energies that are symmetrically distributed around $\omega_p$. A first consequence of the presence of a finite electric field is the broadening of these levels into bands with a width $\sim U_0N_x$. This is clearly visible in Fig.~\ref{fig:1} (c), where we plot the eigenfrequencies $\omega_{\lambda}$ (black dots) as a function of the mean displacement of the corresponding modefunction, $\braket{x} = \sum_i x_i \, |f_{\lambda}(\vec{r}_i )|^2$, for finite $\alpha$ and a finite voltage drop $U_{0}$. Here, the most visible effect of the synthetic electric field is the tilting of the photonic Landau level with a slope proportional to $\sim U_0$. 

In the intermediate magnetic field regime, where $l_0  < l_B < L_{x,y}$ and $l_B=\sqrt{\hbar /eB}=l_0/\sqrt{2\pi \alpha}$ is the magnetic length, Eq.~\eqref{eq:photonic_eigenvalue_eq} can be approximated by a differential equation in the continuum, which recovers the form of a Schr\"odinger equation for a particle in an external electric and magnetic field \cite{DeBernardis_PhysRevLett.126.103603}.
The photonic eigenmodes of the lattice can then be approximated by Landau levels in the continuum, $f_{\lambda}(\vec r_i)\equiv\Phi_{\ell k}(\vec r_i)$, where
\begin{equation}\label{eq:LandauOrbitals}
\Phi_{\ell k}(\vec r) = l_0 \frac{e^{iky}}{\sqrt{L_y}} \varphi_{\ell }^{\rm h.o.} \left( x + l_B^2k - l_B\frac{U_B}{\hbar \omega_B} \right).
\end{equation}
and $\varphi_{\ell }^{\rm h.o.}(x)$ is the $\ell$-th harmonic oscillator eigenfunction with oscillator length given by the magnetic length $l_B$ (see Appendix \ref{app:LandauLevel_E} for more details). 
In Eq.~\eqref{eq:LandauOrbitals}, the index $\ell=0,1,2,\dots $ labels the discrete Landau levels and $k$ is the wavevector along the $y$-direction. We have also introduced the parameter
\begin{equation}
    U_B = eE l_B,
\end{equation}
which characterizes the interplay between the magnetic and the electric field and corresponds to the voltage drop across a cyclotron orbit. In the following we will refer to $U_B$ as the \emph{Landau voltage} and we will see how it plays a crucial role in determining the light-matter coupling dynamics. 

In the presence of the electric field, the Landau levels are no longer degenerate and their energy is approximately given by
\begin{equation}\label{eq:spectrum_LL_w_E}
    \hbar \omega_{\ell k} \approx \hbar\omega_b + \hbar\omega_B \left(\ell+\frac{1}{2}\right) + \hbar \omega_{\ell}^{(2)} + U_B \left( l_B k - \frac{U_B}{2\hbar \omega_B} \right),
\end{equation}
where  $\omega_b=\omega_p-4J$ is the frequency of the lower band edge and  $\omega_B= 4\pi \alpha J$ is the cyclotron frequency.
Here we have also included the second order correction to the Landau levels due to lattice discretization \cite{DeBernardis_PhysRevLett.126.103603},
\begin{equation}
    \omega_{\ell}^{(2)} = -\frac{\omega_B^2}{32J}\left( 2\ell^2 + 2\ell + 1 \right),
\end{equation}
which is necessary to match this analytic result with exact numerics.
In Fig.~\ref{fig:1} (c) we compare Eq.~\eqref{eq:spectrum_LL_w_E} (red dashed lines) to the full spectrum (black dots) obtained via a numerical diagonalization of Eq.~\eqref{eq:photonic_eigenvalue_eq}. 
For the analytic results in this plot, we approximate the average displacement of the eigenmodes by $\braket{x} \simeq - l_B^2 k + U_Bl_B/(\hbar \omega_B)$.
We see that Eq.~\eqref{eq:spectrum_LL_w_E} predicts very accurately the lowest energy bands for the bulk modes, while the highest energy bands are given by a symmetric mirroring of the lowest states. Of course, the continuum approximation fails near the center of the band, where the effect of the discretization become important. As we are going to see in what follows, the linear shape of the dispersion relation in Eq.~\eqref{eq:spectrum_LL_w_E} plays a crucial role since it guarantees that wavepackets do not suffer broadening or distortion during propagation. 

Note that because we consider a finite lattice, we have a limited number of bulk modes in each Landau level $\ell$. The eigenmodes are localized states along the $x$-direction, with a spatial extension of $\Delta x \sim \sqrt{1+\ell} \, l_B$ and centered at the $k$-dependent position $\braket{x} = - l_B^2 k + U_Bl_B/(\hbar \omega_B)$. As such, the number of states in each Landau level can be estimated by counting the number of $k$-modes with spacing $\Delta k=2\pi/L_y$ that cover the distance $L_x$.
For example, based on this estimate, the lowest Landau level (LLL) with $\ell = 0$, contains about $M_{\ell=0} \approx L_xL_y/(2\pi l_B^2) = M\alpha$ states for which $0<k<L_x/l_B^2$. 

The remaining states outside this range are instead fully localized on the edge and represent the usual topological edge modes. In Fig.~\ref{fig:1} (c) we can see that for states at the boundary of the system, where $\braket{x}\sim 0, L_x$, the eigenfrequencies lie outside the tilted Landau level, and form a band of edge states. Note, however, that in contrast to the dispersionless bulk states, the large variation of the group velocity along the edges typically leads to a significant broadening and deformation of propagating wavepackets.

\section{Quantum Hall transport for photons}
\label{sec:photonic_transport_quantumHall}

In this section we provide a theoretical framework to describe the single photon dynamics in this synthetic quantum Hall configuration. In particular, this analysis highlights one of the most important consequences that arises from the presence of both magnetic and electric fields: the photonic bulk, often referred as a Chern insulator, allows now for propagation and transport in a direction perpendicular to the two fields.
This is the direct photonic analog of the quantum Hall effect for electrons. 

\subsection{Photon Green's function}

In order to describe the quantum Hall dynamics we make use of the photonic Green's function (or propagator).
In its general form, this Green's function can be expressed in terms of the photonic eigenmodes and their corresponding eigenfrequencies as
\begin{equation}
    \begin{split}
    G(t, \vec{r}_i, \vec{r}_j ) & = \braket{ {\rm vac}| \Psi (t, \vec{r}_i) \Psi^{\dag}(0, \vec{r}_j) | {\rm vac} } 
    \\
    &= \sum_{\lambda}f_{\lambda}(\vec{r}_i)f^*_{\lambda}(\vec{r}_j)e^{- i \omega_{\lambda} t}
    \end{split}
\end{equation}
and provides the propagation amplitude of a photon from position $\vec r_j$ to position $\vec r_i $ in a time $t$. 

Using the approximated forms of the photonic eigenmodes and eigenfrequencies given in Eqs.~\eqref{eq:LandauOrbitals}-\eqref{eq:spectrum_LL_w_E}, we can  write down an explicit expression for the Green's function restricted to the LLL,
\begin{equation}\label{eq:Landau_Green's_fun_UB}
    \begin{split}
        G_{\ell=0}(t, \vec{r}_i, \vec{r}_j )  \approx & \frac{l_0^2}{2\pi l_B^2} e^{i(\theta_{ij}-\omega^{LL}_{\ell=0} t)}e^{- \frac{|\vec{r}_i - \vec{r}_j|^2}{4 l_B^2}}  I(t, \vec{r}_i, \vec{r}_j).
    \end{split}
\end{equation}
Here $\theta_{ij}$ is the (gauge-dependent) phase
\begin{equation}
    \theta_{ij} = -\frac{x_iy_j - x_j y_i}{2l_B^2} + \frac{x_i y_i - x_j y_j}{2l_B^2},
\end{equation}
and 
\begin{equation}
\label{eq:zp}
   \omega_{\ell}^{LL} = \omega_b + \omega_B\left( \ell + \frac{1}{2}\right) + \omega_{\ell}^{(2)}
\end{equation}
is the frequency of the $\ell$-th Landau level.
%zero-point frequency. 
%\pr{(sign of $U_B^2$ term correct? Find better name for this frequency)} 

The electric field enters in the dynamics only through the time-dependent part of the propagator, $I(t, \vec{r}_i, \vec{r}_j)$, which is given by
\begin{equation}\label{eq:time-dep_propagator}
\begin{split}
    I(t, \vec{r}_i, \vec{r}_j) = & \frac{2\sqrt{\pi} l_B}{L_y} \sum_{k} e^{ - \left( l_B k - \frac{(x_i+x_j) - i (y_i -y_j )}{2l_B} + \frac{U_B}{\hbar \omega_B} \right)^2 } 
    \\
    & \times e^{i \left[c_H k + U_B^2/(2\hbar^2\omega_B) \right] t}.
    \end{split}
\end{equation}
Here the Hall speed of propagation is given by the usual formula
\begin{equation}\label{eq:Hall_speed}
    c_H = \frac{U_B l_B}{\hbar} = \frac{E}{B}.
\end{equation}

While there is no simple compact form for Eq.~\eqref{eq:time-dep_propagator} in the general case, it is possible to simplify the problem in the limit $L_y \rightarrow \infty$. In this limit, the sum in Eq.~\eqref{eq:time-dep_propagator} can be approximated by an integral by substituting $2\pi/L_y \sum_k \mapsto \int dk$, and we obtain
\begin{equation}\label{eq:time-dep_propagator_infinite}
    I(t, \vec{r}_i, \vec{r}_j) \simeq e^{- \frac{U_B^2t^2}{4 \hbar^2}} e^{i  \left( \frac{x_i + x_j + i (y_i - y_j)}{2l_B} - \frac{U_B}{2\omega_B}\right) U_Bt/\hbar}\,.
\end{equation}
For a vanishing magnetic field, $U_B = 0$, this expression reduces to a constant, $I(t, \vec{r}_i, \vec{r}_j)|_{U_B = 0} = 1$, meaning that photons do not propagate. In the presence of emitters, this property gives rise to the formation of localized Landau-photon polaritons, as described in Ref. \cite{DeBernardis_PhysRevLett.126.103603}.

By combining Eq.~\eqref{eq:time-dep_propagator_infinite} and Eq.~\eqref{eq:Landau_Green's_fun_UB} we obtain the total Green's function in the continuum for an infinitely large system,
\begin{equation}\label{eq:Green's_fun_PBC_infinitey_equalx}
    \begin{split}
        G_0(t, \Delta x, \Delta y ) & = \frac{l_0^2}{2\pi l_B^2}e^{-\frac{\Delta x^2}{4 l_B^2}}e^{-\frac{1}{4}\left( U_B t/\hbar - \Delta y/l_B \right)^2 } 
        \\
        & \times e^{i\left[ \theta_{ij} - \left(\omega_{\rm ch}(x_j) + \omega_{\rm ch}(x_i) \right)t/2 \right]}.
    \end{split}
\end{equation}
Here, $\Delta x = x_j - x_i$ and  $\Delta y= y_j -y_i$, and 
%where 
we have introduced the position-dependent Hall-channel resonance frequency
\begin{equation}\label{eq:zero-point-freq-space-dep}
    \omega_{\rm ch}(x) = \omega_{\ell=0}^{LL} - \frac{U_B}{\hbar} \frac{x}{l_B} + \frac{U_B^2}{2\hbar^2 \omega_B}. 
\end{equation}
which includes a second order correction due to mixing with the higher Landau levels by the synthetic electric field.
From Eq.~\eqref{eq:Green's_fun_PBC_infinitey_equalx} we see that for $\Delta y > 0$ the photon emitted in $y_j$ can coherently propagate to $y_i$ at the Hall speed $c_H$, provided that $|x_i-x_j|\lesssim l_B$. On the other hand, for $\Delta y < 0$ the propagation is exponentially suppressed (for $t>0$). This clearly shows that the photon propagation is unidirectional (or chiral) and without any dispersion.

In summary, our calculations show that for each point in the bulk, the photonic lattice behaves as a unidirectional waveguide along the $y$-direction, perpendicular to both the electric and the magnetic field and with a Gaussian transverse size $\Delta x \sim l_B$ fixed by the magnetic length. This is  schematically shown in Fig.~\ref{fig:1} (a).
Each chiral channel at position $x$ has its own resonance frequency $\omega_{\rm ch}(x)$, and it is detuned from neighboring channels at positions $x\pm l_0$ by $\hbar |\Delta \omega| = U_0$.

{We emphasize that these considerations are only valid in the regime of weak magnetic field strengths \cite{DeBernardis_PhysRevLett.126.103603}, which is the regime of interest for the current study and holds approximately for $\alpha \lesssim 1/6$. For larger values of the magnetic flux the dynamics is more complicated and exhibits Bloch oscillations and other discrete-lattice effects. For the sake of clarity we do not address this regime here and refer to Refs.~\cite{Price_PhysRevA.85.033620, Yongguan_PhysRevB.104.104314} for further details. In any case, note that all the presented results have been benchmarked against numerical simulations of the full lattice dynamics, i.e., with no approximations.}

\subsection{Local density of states}

The photonic Green's function derived above is useful not only to describe the propagation of a photon through the lattice, but also to extract the local density of states, defined as
\begin{equation}\label{eq:def_local_DOS}
    \rho_{\rm ph}(\vec{r}, \omega) = \int d t \, G(t, \vec{r}, \vec{r} ) e^{i \omega t}.
\end{equation}
In the following we will show that this quantity is particularly important for describing the dynamics of a single emitter that is coupled to the photonic lattice at a position $\vec r$.

From Eq.~\eqref{eq:Green's_fun_PBC_infinitey_equalx} we can derive an analytic expression for the local density of states for the LLL,
\begin{equation}\label{eq:DOS_ph_approximate}
    \rho^{\ell=0}_{\rm ph}(\vec{r}, \omega) \approx \frac{2\sqrt{\pi}\alpha \hbar}{U_B}
    %e^{-\frac{(\hbar \omega)^2}
    e^{-\frac{\hbar^2\left[\omega-\omega_{\rm ch}(x)\right]^2}     {U_B^2}}\,.
\end{equation}
This expression shows that as long as the relevant system dynamics takes place in a narrow range of frequencies $\hbar |\omega-\omega_{\rm ch}(x)| \lesssim U_B$, the bulk behaves effectively as a continuous 1D waveguide, with an almost constant density of states. However, for a lower value of the electric field or when the dynamics involves a wider range of frequencies, $\hbar |\omega-\omega_{\rm ch}(x)| > U_B$, the density of states decays very rapidly and non-Markovian effects start to play a relevant role. In this regime, the direct analogy with a conventional chiral waveguide breaks down and new phenomena appear, as we are going to see in the next Section.

The shape of the density of states in Eq.~\eqref{eq:DOS_ph_approximate} is quite surprising at first sight. Indeed, by looking at Fig.~\ref{fig:1} (c) and at the analytic estimates given above, one would expect an approximately flat density of states over the whole width $\sim U_{0}M_x$ of the tilted Landau level, where $M_x=L_x/l_0$. Instead, our calculations show that the effective photonic bandwidth within each Landau level that is accessible to a localized emitter is determined by the Landau voltage $U_B$, independently of the lattice size. This is due to the peculiar lateral localization of the chiral bulk modes at a $k_y$-dependent $x$ position, which modulates the effective light-matter coupling.

Before we proceed, it is important to emphasize how this behaviour is different from the one of standard systems in chiral quantum optics. A non-trivial shape of the photonic density of states is in fact also seen by an emitter coupled to the chiral edge states, the frequency variation arising from the frequency-dependent penetration length of the edge states into the bulk. In contrast to our case, however, this density of states %%has a more complex structure and, in particular, 
is relatively smooth and does not display a quick Gaussian decay away from the central frequency. This seemingly minor difference is at the heart of the state-transfer application that we are going to discuss for the bulk modes in the next sections.

%\section{Single emitter interactions in the synthetic quantum Hall regime}
\section{Coupling regimes in photonic quantum Hall systems}
\label{sec:3_regimes_LM_int}

Let us now go beyond the sole propagation of photons and consider an additional quantum emitter located at a position $\vec r_e=(x_e,y_e)$ in the bulk of the lattice, as described by the total Hamiltonian in Eq. \eqref{eq:ham_total_light-matter}, with $N=1$.
Since there is only a single emitter here, we suppress the index $n$ and set $\omega_e^n, g_n \mapsto \omega_e, g$.

By assuming that the emitter is initially prepared in its excited state with no photons in the lattice, the resulting dynamics of the system is constrained to the single excitation subspace and can be described by a wavefunction of the form  
\begin{equation}\label{eq:StateAnsatz}
|\psi\rangle(t)= e^{-i\omega_{e} t}[c_e(t) \sigma_+ + \sum_i \varphi(\vec r_i,t) \Psi^\dag(\vec r_i)]|g\rangle|{\rm vac}\rangle\,.
\end{equation}
By projecting the Schr\"odinger equation onto this subspace, we can derive a set of equations for the time evolution of the emitter amplitude $c_e(t)$ and for  the wavefunction of the photon, $\varphi(\vec{r}_i,t)$~\cite{Chang_RevModPhys.90.031002, Fang_2018, Dinc2019exactmarkoviannon}. 
These equations can be readily integrated numerically, which we use to produce most of the results discussed in the following. Alternatively, we can eliminate the dynamics of the emitted photon to derive a closed equation for the emitter amplitude \cite{DeBernardis_PhysRevLett.126.103603}, 
\begin{equation}\label{eq:eq_retarded_single_atom}
\dot{c}_e(t) = - \frac{g^2}{4}  \int_0^t ds \,G(t-s,\vec r_e, \vec r_e) c_e(s) e^{i\omega_e(t-s)}.
\end{equation} 
This is an integro-differential equation, with the photonic Green's function evaluated at the emitter position $\vec{r}_e$ as the memory kernel. This memory kernel describes both the photon's emission from the atom and its eventual re-absorption. 

While in general there is no closed analytic solution of Eq. \eqref{eq:eq_retarded_single_atom}, we can use an approximate expression for the photonic Green's function to obtain additional useful insights into the emitter-photon dynamics. In particular, with the help of the Gaussian approximation for the LLL in Eq. \eqref{eq:DOS_ph_approximate}, we obtain
\begin{equation}\label{eq:eq_retarded_single_atom_GaussianApprox}
    \dot c_e(t) = -\frac{g^2 \alpha}{4} \int_0^t ds \, e^{-U_B^2(t-s)^2/(4\hbar^2)} \, e^{i \Delta_e (t-s)}\,c_e(s)\,,
\end{equation}
where we have introduced the position-dependent detuning 
\begin{equation}
    \Delta_e = \omega_e - \omega_{\rm ch} (x_e).
\end{equation}
Under the assumption $\Delta_e\approx0$, this approximate form allows us to identify three qualitatively different coupling regimes:
\begin{enumerate}
    \item \emph{Weak-coupling} regime, $\hbar g\sqrt{\alpha} \ll U_B$. In this limit the density of states is almost flat and the Green's function decays on a timescale that is fast compared to the evolution of the emitter. This leads to an effectively Markovian dynamics, with an exponential decay of the excited state. 
    
    \item \emph{Strong-coupling} regime, $\hbar g\sqrt{\alpha} \gg U_B$. Under this condition the coupling strength exceeds the relevant bandwidth $U_B$ of the density of states and the emitted photons can be reabsorbed before they propagate away. Such conditions lead to the formation of so-called atom-photon bound states \cite{Calajo_PhysRevA.93.033833,Cirac_PhysRevX.6.021027}, which  do not decay.   
    
    \item \emph{Critical-coupling} regime, $\hbar g\sqrt{\alpha} \simeq U_B$. For these parameter values, the absence of a sharp band-edge allows the excited state population to fully decay to zero, but the sizable frequency-dependence of  the  density of states makes a crucial difference from other narrow-band waveguide systems and, as we are going to discuss in detail below, this results in a strongly non-Markovian  dynamics and, in particular, in an almost symmetric shape of the spontaneously emitted photon wavepacket.  
\end{enumerate}
Note that the relevant coupling parameter in this discussion is $g\sqrt{\alpha}$, rather than the bare light-matter coupling $g$. This is related to the fact that the local density of states scales as $\sim \alpha$, resulting in an additional factor $\sqrt{\alpha}$ in the effective coupling strength \cite{DeBernardis_PhysRevLett.126.103603}. Physically, this factor can also be understood from the spatial width $l_B\sim 1/\sqrt{\alpha}$ of the waveguide mode in the transverse direction.

In the following we proceed with a brief discussion of the photon-emission dynamics in those three regimes for a situation, where the emitter is far away from the boundary. 

\subsection{Weak coupling regime: Markovian spontaneous emission}
The linear dispersion of the photonic Landau levels introduced by the electric field $E$ and captured by Eq. \eqref{eq:spectrum_LL_w_E} implies that photons can propagate across the lattice. Therefore, a photon locally created by the emitter can leave the interaction region, which leads to spontaneous decay. This is in stark contrast to the case $E=0$, where spontaneous emission is forbidden and is replaced by the formation of bound polaritonic states between the emitter and the localized Landau photons~\cite{DeBernardis_PhysRevLett.126.103603,Leonforte_PhysRevLett.126.063601}. 

In the standard theory of spontaneous decay \cite{milonni1994quantum}, the exponential decay of the atomic population arises from a Markov approximation. The validity of this approximation requires that the density of states of the photonic is approximately constant over a sufficiently large frequency range. In the current setting, this is the case when the relevant timescale of the emitter's dynamics given by $g\sqrt{\alpha}$ is slow compared to the inverse of the effective bandwidth $U_B/\hbar$ identified above. Under this assumption we can approximate the photonic Green's function appearing in Eq. \eqref{eq:eq_retarded_single_atom} as
\begin{equation}
    G(t-t', \vec{r}, \vec{r}) \approx \rho_{\rm ph}(\vec{r}, \omega_e ) \, \delta (t-t'),
\end{equation}
and obtain a Markovian, i.e., memoryless equation for the emitter amplitude
\begin{equation}
    \dot c_e (t) \approx - \frac{g^2}{4} \rho_{\rm ph}(\vec{r}_e, \omega_e ) \, c_e(t).
\end{equation}
By using the continuum approximation for the Green's function in Eq.~\eqref{eq:DOS_ph_approximate}, we find that the excited state amplitude decays exponentially,
\begin{equation}\label{eq:exp_decay}
    c_e(t) \approx e^{- \Gamma_e t},
\end{equation}
where the decay rate
\begin{equation}\label{eq:decay_rate_GammE}
    \Gamma_e = \frac{\sqrt{\pi}}{2}\frac{\hbar g^2\alpha }{U_B} e^{-\hbar^2\Delta_e^2/U_B^2}
\end{equation}
is inversely proportional to the applied electric field $U_B\sim E$ and has the expected Gaussian dependence on the position-dependent emitter detuning $\Delta_e$.

\begin{figure}
    \centering
    \includegraphics[width=\columnwidth]{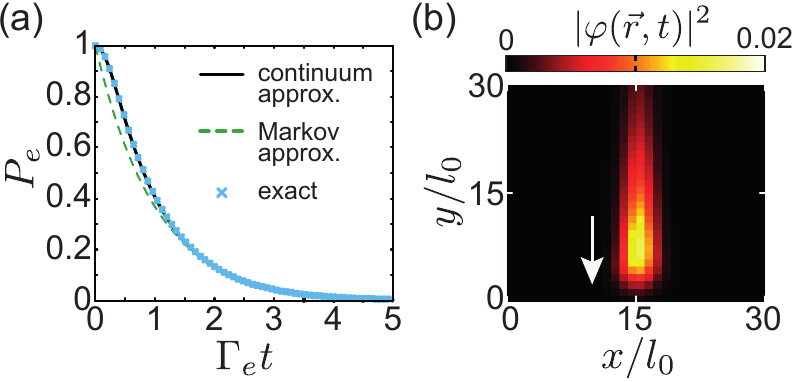}
    \caption{Spontaneous emission dynamics of a single emitter in the weak-coupling regime. (a) Plot of the excited-state population $P_e = |c_e(t)|^2$ as a function of time. 
    The blue crosses represent the results from an exact numerical simulation, while the solid black line shows the prediction from Eq. \eqref{eq:eq_retarded_single_atom_GaussianApprox}. The green dashed line indicates an exponential decay with a rate $\Gamma_e$ given in Eq. \eqref{eq:decay_rate_GammE}.  (b) Snapshot of the photon density $|\varphi (\vec{r},t)|^2$ of the emitted wavepacket at a time $t>0$ well after the spontaneous emission process. The white arrow marks the direction of propagation. The parameters for both plots are $\alpha = 1/20$, $N_x=31$, $N_y=100$, $U_0 /J = 0.001$, $\Delta_e/J \approx 0$ and $\hbar g = 0.4 U_B/\sqrt{\alpha} \approx 0.003 \hbar J$. The emitter is located at the position $\vec{r}_e/l_0 = (15, 50)$.}
    \label{fig:2}
\end{figure}

In Fig.~\ref{fig:2} (a) we perform an exact numerical simulation of the decay of an excited emitter in a finite lattice with periodic boundary conditions along the $y$-direction. These results are compared to the dynamics predicted by Eq.~\eqref{eq:eq_retarded_single_atom_GaussianApprox} for the continuum limit. We see that in this weak-coupling regime, the continuum approximation is in excellent agreement with exact results. This simulation also confirms that apart from small deviations at the initial stage, the temporal profile of the
decay is very well captured by an exponential decay with a rate $\Gamma_e$ given in Eq.~\eqref{eq:decay_rate_GammE}.

In Fig.~\ref{fig:2} (b) we also show a snapshot of the emitted photonic wavepacket, $|\varphi(\vec{r},t) |^2$. This plot confirms that the emitted photon is localized along the $x$-axis within a magnetic length $l_B$ and propagates along the $y$-axis with Hall speed $c_H$.
As it is typical for spontaneous emission, the wavepacket is asymmetrically stretched along the propagation direction, with a sharp front edge and a long exponential tail of characteristic length $\sim c_H/\Gamma_e$.

\subsection{Strong coupling regime: Atom-photon bound states in the chiral continuum}

For very large values of the light-matter coupling, $\hbar g > U_B/\sqrt{\alpha}$, the emitter dynamics is dominated by non-Markovian effects, which arise from the finite width of the density of states given in Eq.~\eqref{eq:DOS_ph_approximate}.
In particular, the density of states is strongly suppressed outside a frequency band of width $\sim U_B$, meaning that the Green's function in Eq.~\eqref{eq:eq_retarded_single_atom} can be approximated by a constant, $G(t-s,\vec r_e)\approx \alpha$, over the relevant timescale of the emitter dynamics. This allows us to approximate the equation for the emitter amplitude $c_e(t)$ by a second-order differential equation.
\begin{equation}
    \ddot c_e(t) \approx - \frac{\Omega^2}{4} c_e(t),
\end{equation}
where the Rabi frequency is given by
\begin{equation}
    \Omega = g\sqrt{\alpha}.
\end{equation}
This equation indicates the presence of an atom-photon bound-state \cite{Calajo_PhysRevA.93.033833, Cirac_PhysRevX.6.021027}, as an exact eigenstate of the system, and recovers the Landau-photon polariton (LPP) picture described in \cite{DeBernardis_PhysRevLett.126.103603}. 
Indeed, numerical simulations confirm that the photonic component of this bound state has the shape of a standard Landau orbital with no visible effect from the electric field. 
However, while for $E=0$ LPPs already appear at arbitrarily small values of the coupling strength~\cite{DeBernardis_PhysRevLett.126.103603}, in the present configuration bound states require a minimal coupling strength that exceeds $U_B$. 
%However, compared to the formation of those LPPs, which in the case of $E=0$ already appear at arbitrarily small coupling strength, the formation of bound states in quantum Hall systems requires a minimal coupling strength that exceeds $U_B$.

\begin{figure}
    \centering
    \includegraphics[width=\columnwidth]{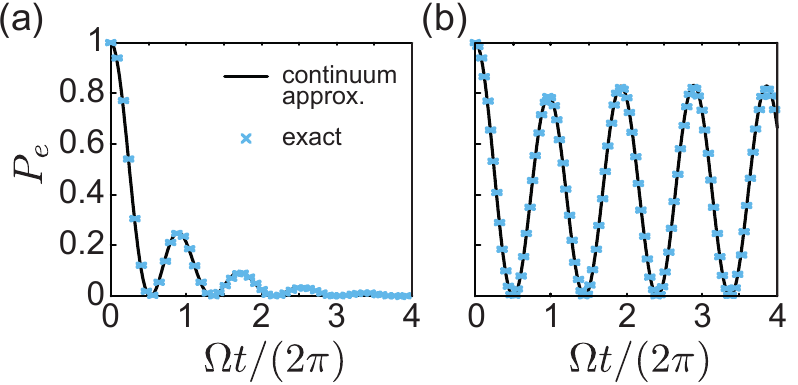}
    \caption{ Evolution of the excited state population $P_e = |c_e(t)|^2$ of an initially excited emitter in the strong-coupling regime, where (a) $\hbar g=2 U_B/\sqrt{\alpha}$ and (b) $\hbar g=8 U_B/\sqrt{\alpha}$. The blue crosses represent the results from an exact numerical simulation, while the solid black line shows the prediction from Eq. \eqref{eq:eq_retarded_single_atom_GaussianApprox} in the continuum limit. The other parameters assumed for both plots are $\alpha = 1/20$, $N_x=N_y=40$, $U_0 /J = 0.01$ and $\Delta_e/J \approx 0$.}
    \label{fig:3}
\end{figure}

In Fig.~\ref{fig:3} (a) and Fig.~\ref{fig:3} (b) we show the emitter dynamics as we progressively enter the strong coupling regime. Already for $\hbar g\sqrt{\alpha}=2 U_B$ we see clear non-Markovian effects and marked oscillations, but at longer times the emitter dynamics keeps being dominated by a monotonous decay. 
For higher coupling strengths, $\hbar g\sqrt{\alpha}=8 U_B$, we only observe a small initial decay followed by persistent Rabi oscillations between the photonic and the matter components of the bound state. 
The fact that the Rabi oscillations in Fig. ~\ref{fig:3} (b) are incomplete, i.e., $P_e(t=2\pi n/\Omega)< 1$ for $n=1,2\ldots$, is due to the smoothly decaying tail of the density of states, which still allows a small fraction of the excitation to propagate away into the lattice. By further increasing the light-matter coupling $\hbar g \sqrt{\alpha} \gg U_B$ the Rabi oscillations progressively reach their maximum value $P_e(t=2\pi n/\Omega)\approx 1$.
These observations are consistent with the formation of atom-photon bound states in other narrow-band waveguide QED systems. 

However, we re-emphazise that in our case this effect appears under conditions where the effective coupling is still small compared to the total width of the LLL and, in general, the total lattice band-width, $\hbar g \sqrt{\alpha} \ll U_0 N_x\ll 8\hbar J$.
{Interestingly, note that in an infinite lattice the photon has a continuum spectrum and, strictly speaking, these atom-photon bound states belong to the class of so-called \emph{bound state in the continuum} (BIC) \cite{BIC_review, Calajo_PhysRevLett.122.073601}. This may look quite odd, since BICs are thought not to exist in geometries where photons propagate chirally. The controversy is easily solved by noticing that only the local density of states matters for this kind of bound states. As it was previously highlighted, even though the photon has a continuum spectrum, the states accessible to the atom are limited by the Gaussian density profile Eq. \eqref{eq:DOS_ph_approximate}. This makes the photon to locally have an effectively finite bandwidth, and thus to follow the standard rules for atom-photon bound states in conventional finite-band configurations~\cite{Calajo_PhysRevA.93.033833, Cirac_PhysRevX.6.021027}.
}

\subsection{The critical-coupling regime}

\begin{figure}
    \centering
    \includegraphics[width=\columnwidth]{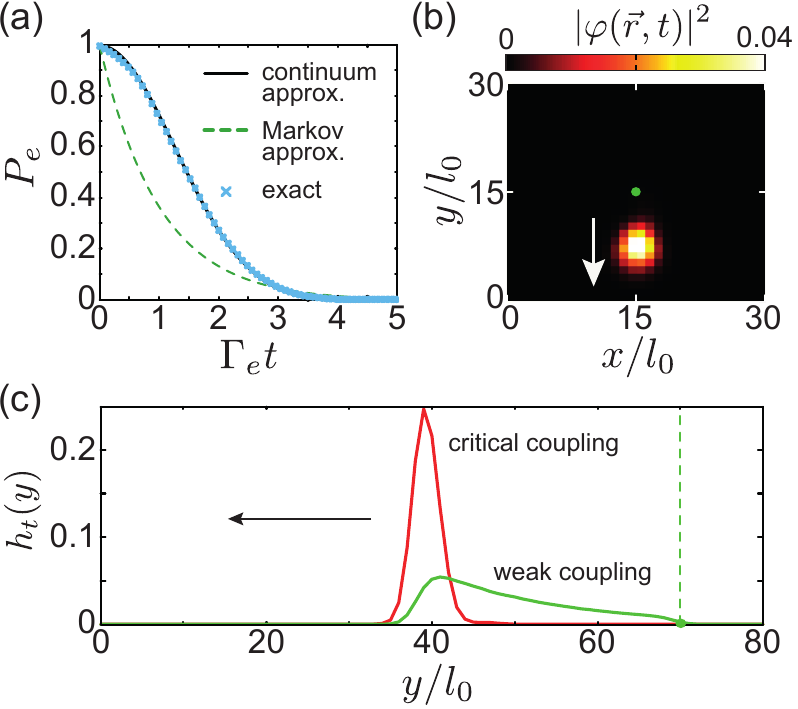}
    \caption{(a) Excited state population of a single emitter $P_e = |c_e(t)|^2$ as a function of time. The blue crosses represent the results from an exact numerical simulation, while the solid black line shows the prediction from Eq. \eqref{eq:eq_retarded_single_atom_GaussianApprox} in the continuum limit. The green dashed line indicates an exponential decay with a rate $\Gamma_e$ given in Eq. \eqref{eq:decay_rate_GammE}.  
     (b) A snapshot of the photon density $|\varphi (\vec{r},t)|^2$ of the emitted wavepacket at a time $t\approx 202 J^{-1} \approx 23 \Gamma_e^{-1}$. The white arrow marks the direction of propagation.
    In both plots we assume $\alpha = 1/10$, $N_x=N_y=31$, $U_0 /(\hbar J) = 0.1$, $\Delta_e/J = 0$ and $\hbar g = U_B/\sqrt{\alpha} \approx 0.4 \hbar J$. The emitter is located at $\vec{r}_e/l_0=(15,15)$, as indicated by the green dot.
    (c) Integrated photon density profile $h_t (y)=\int dx |\varphi(x,y,t)|^2$ for a photon emitted in the Markovian regime with $\hbar g = 0.3 U_B/\sqrt{\alpha}$ (green line) and from a critically coupled emitter with $\hbar g = U_B/\sqrt{\alpha}$ (red line). The black arrow indicates the propagation direction while the green dashed line marks the emitter position at $\vec{r}_e/l_0 = (10, 70)$. The two snapshots are taken at the same time $Jt\approx 64$. In this plot we assume $\alpha=1/10$, $N_x=20$, $N_y=80$, $U_0/(\hbar J) = 0.1$ and $\Delta_e/J = 0$. Both results in (c) have been obtained from a numerical simulation of the full lattice dynamics.}
    \label{fig:4plus}
\end{figure}

As we increase the coupling strength from the weak to the strong coupling regime, the system passes through a critically coupled regime where 
\begin{equation}
    \hbar g \approx \frac{U_B}{\sqrt{\alpha}}\,.
\end{equation}
As shown in Fig.~\ref{fig:4plus} (a), under this condition the decay dynamics of the excited emitter is no longer Markovian, but Rabi oscillations, as characteristic for the strong-coupling regime, are not yet visible either. This condition is of interest for two reasons. First of all, setting $\hbar g\approx U_B/\sqrt{\alpha}$ is the largest coupling that still allows spontaneous emission before being suppressed in the strong-coupling regime, resulting in the fastest way to fully de-excite the emitter in a time $\Gamma_e^{-1}\approx U_B^{-1}$. Secondly, when looking at the shape of the emitted photon in Fig.~\ref{fig:4plus} (b), we find that the wavepacket is very compact, also along the $y$-direction. This property is shown in more detail in Fig.~\ref{fig:4plus} (c), where we plot a cut of the emitted photon wavefunction along the $y$-direction and we compare it with the one obtained in the weak-coupling regime: we see that in the critical coupling case the wavepacket is not only highly localized, but also almost symmetric around its maximum and travels through the lattice without any significant dispersion. 

While the formation of effective 1D chiral channels is a property of the photonic lattice itself, the emission of such symmetric wavepackets is connected to a specific light-matter interaction regime and goes beyond the usual quantum Hall physics. While this seems to be a minor detail for the emission process, in what follows we will show how this symmetry becomes an essential property when studying the reabsorption of the photon by other emitters in the system. 

\subsection{Beyond the single Landau level approximation}

Our discussion of the different coupling regimes  was so far based on the assumption that the emitter is primarily coupled to the states in the LLL.  This assumption is justified as long as the light-matter coupling $g$ % \ll \Delta_{\rm gap}$ 
is small compared to the gap $\Delta_{\rm gap} \sim \omega_B$ between the Landau levels. This corresponds to $g \ll \omega_B$. In order to reach the critical or strong coupling conditions, we require $\hbar g \gg U_B/\sqrt{\alpha}$, such that $U_B/(\hbar \omega_B) = U_0/(2(2\pi \alpha)^{3/2}J)$ can be of order $\sim O(1)$ already for moderately strong electric fields. This means that the restriction to the LLL might not be well justified in this regime.

\begin{figure}
    \centering
    \includegraphics[width=\columnwidth]{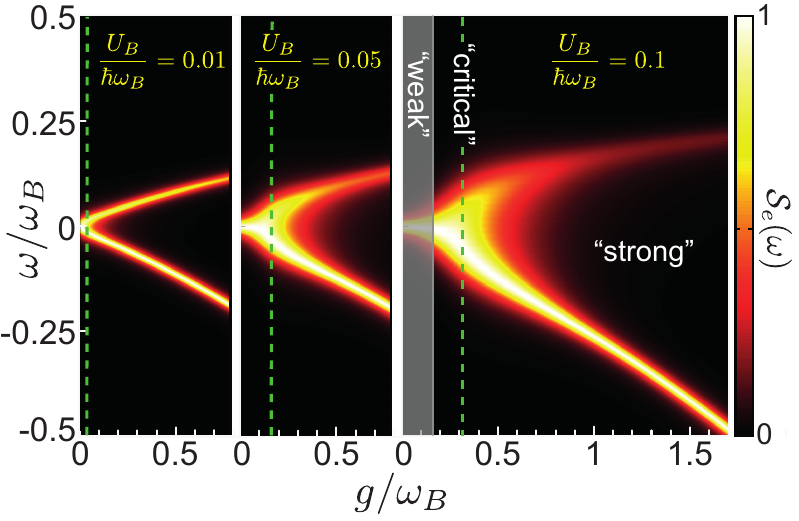}
    \caption{Plot of the emitter excitation spectrum $\mathcal{S}_e$ as a function of the excitation frequency $\omega$ and the light-matter coupling strength $g$. The three panels show this spectrum for different values of the electric field, expressed in terms of the Landau voltage $U_B= eE l_B$.  
    In each panel, the green dashed line marks the critical coupling value $\hbar g =U_B/\sqrt{\alpha}$. The grey-shaded area in the last panel is the weak-coupling (Markovian) regime, whose boundary is given by $\hbar g \approx U_B/(2\sqrt{2\alpha})$, as it is defined in traditional cavity QED setups with a Gaussian inhomogeneous broadening of the emitters~\cite{Molmer_PhysRevA.83.053852, Diniz_PhysRevA.84.063810,DeBernardis_PhysRevB.106.224206}.
    The other parameters are $\alpha = 1/10$, $N_x = N_y = 30$ and  $\Delta_e/J \approx 0$. In all plots an artificial broadening $\gamma_{\delta}/J = 0.015$ of all states has been introduced to obtain a smooth spectrum.}
    \label{fig:4}
\end{figure}

To investigate the influence of higher Landau levels, we analyze the excitation spectrum of the emitter,
\begin{equation}
    \mathcal{S}_e (\omega) = \sum_{\nu} |\braket{\nu|\sigma_+^e|G}|^2\delta ( \omega - \omega_{\nu} ).
\end{equation} 
Here $|\nu \rangle$ and $\omega_{\nu}$ are the $\nu$-th eigenstate and eigenfrequency of the full coupled Hamiltonian in Eq. \eqref{eq:ham_total_light-matter}, while $|G\rangle$ is its ground state. 

The excitation spectrum $\mathcal{S}_e (\omega) $ is plotted in Fig. \ref{fig:4} as a function of the coupling strength $g$ and for different strengths of the electric field. 
In the strong coupling regime, where $\hbar g \gg U_B/\sqrt{\alpha}$, $\mathcal{S}_e (\omega)$ displays two branches which are split by $\Omega$ and corresponds to the bound states discussed above.
For very small electric fields, i.e., $U_B \ll \hbar \omega_B$, 
the two branches are very narrow and almost symmetric, with only a small downward shift of the frequency of the upper bound state. 
This shift of the upper branch can be understood as a second-order correction  given by the presence of the higher Landau level. However, since
the states in the $\ell =1$ Landau level are spatially well separated from the $\ell =0$ states at the same energy, as one can see in Fig.~\ref{fig:1} (c), their effect is still small, acting only perturbatively on the bound-state dynamics.
%This can be understood by looking at the spectrum shown in Fig.~\ref{fig:1} (c). 
%The only effect of the higher Landau level is a slight, second-order downward shift of the frequency of the upper bound state. 

By increasing the electric field to values of $U_B \approx \hbar \omega_B$,  the coupling to the higher Landau level becomes more relevant, since resonant states in different $\ell$-manifolds have a larger spatial overlap. This is most visible for the upper bound-state. As its energy approaches the next Landau level, it becomes progressively more broadened, due to the increasing possibility to decay into propagating modes in the $\ell=1$ manifold. The effect on the lower bound-state is much weaker as this state is further detuned from the $\ell=1$ levels and thus the effective tunneling barrier to resonant propagating states is wider.

More quantitatively, the higher Landau levels result in additional peaks in the density of states, which are separated by multiples of $\hbar\omega_B$ and can be well approximated by
\begin{equation}\label{eq:DOS_higher_LL}
    \begin{split}
        \rho^\ell_{\rm ph}(\vec{r}, \omega)  \approx  &\frac{2\sqrt{\pi}\alpha \hbar}{\sqrt{2^{\ell} \, \ell !} \,U_B} H_{\ell}^2\left(\omega - \omega_{\rm ch}(x)-\ell\omega_B \right)
        \\
        &  \times e^{-\frac{\hbar^2\left[ \omega-\omega_{\rm ch}(x)-\ell\omega_B\right]^2}{U_B^2}}, 
    \end{split}
\end{equation}
where $H_{\ell}(x)$ is the $\ell$-th Hermite polynomial. From this approximate expression for the density of states, 
we can interpret higher Landau levels in the presence of an electric field as regular Landau levels that are shifted in space by $\Delta x_B \approx \ell \hbar \omega_B/U_0$. In this picture, the negligible coupling to neighboring Landau levels can be explained in terms of the reduced spatial overlap $\sim\exp(-\ell \omega_B^2/U_B^2)$ between the wavefunctions, which is strongly suppressed, unless the electric field is very strong. 

Finally, for very strong electric fields, $U_B \gtrsim \hbar \omega_B$, as shown in the right panel of Fig.~\ref{fig:4}, the strong-coupling regime cannot be reached without having the light-matter coupling comparable or even larger than the energy gap between the neighboring Landau levels. As a consequence, the upper bound-state is strongly broadened by a significant hybridization with a wide band of  propagating states in the $\ell =1$ Landau level.

%\section{Population revivals and state-transfer in the critical-coupling regime}
\section{Quantum revivals and state transfer}
\label{sec:periodic_rev_ballistic}

In this section we now explore in more detail one of the most remarkable features of the critical coupling regime, namely the symmetry between emission and absorption processes. Due to the symmetric shape of the emitted wavepacket and its unidirectional propagation, the emission of a photon in this regime is indistinguishable from the time-reversed reabsorption process of the same photon. In the quantum communication literature \cite{Cirac_PhysRevLett.78.3221}, this symmetry argument has been used to derive specific control pulses $g(t)$, which produce such symmetric wavepackets and thus allow for high-fidelity state transfer operations in unidirectional Markovian channels. Here we find that in our proposed configuration this symmetry emerges naturally and without any time-dependent control from the non-Markovian dynamics of a critically coupled photonic quantum Hall system.      

\begin{figure}
    \centering
    \includegraphics[width=\columnwidth]{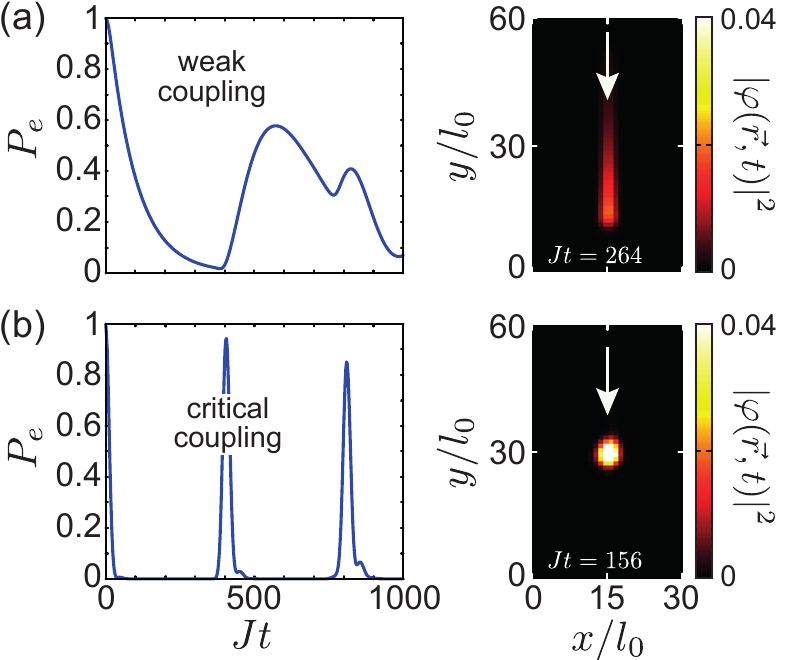}
    \caption{Evolution of the excited state population  $P_e = |c_e(t)|^2$ (left panels) and snapshot of the emitted photon density $|\varphi (\vec{r},t)|^2$ (right panels). In (a) the emitter is weakly coupled with $g\sqrt{\alpha} =  0.3 \, U_B/\hbar \approx 0.038\,J$, while (b) shows the case of a critically coupled emitter with $g\sqrt{\alpha} = U_B/\hbar \approx 0.126\, J$. In both cases $J\tau_{\rm rev} \approx 390$.  The other parameters for these plots are $\alpha = 1/10$, $N_x = 31$, $N_y = 61$, $U_0 /(\hbar J) = 0.1$, $\Delta_e = 0$ and the emitter is located at $\vec{r}_e/l_0 = (15, 53)$. All results have been obtained from a numerical simulation of the full lattice dynamics.}
    \label{fig:5}
\end{figure}

\subsection{Quantum revivals}\label{subsec:revivals}

To analyze reabsorption processes in our system, let us stick to the case of a single emitter, but now in a lattice with periodic boundary conditions (PBC) along the $y$-direction. In this case the emitted photon still propagates unidirectionally with a group velocity set by the Hall speed $c_H$ and without any significant dispersion. However, after a round-trip time
\begin{equation}\label{eq:rev_time}
    \tau_{\rm rev} = L_y/c_H
\end{equation}
the photon will reach again its initial position, where it can be partially or fully reabsorbed by the emitter. 

In Fig.~\ref{fig:5} (a) and (b) we simulate this emission and reabsorption process under weak-coupling and critical-coupling conditions, respectively. In the first case, we see that after each round-trip only a fraction of about $60\%$ of the initial excitation is reabsorbed. This is very similar to what is expected for photon reabsorption in a 1D Markovian channel without any time-dependent control. In contrast, for a critically coupled emitter, the photon is reabsorbed with more than $P=95\%$ probability, and significant revivals can still be observed after multiple roundtrips. This near perfect reabsorption can also be interpreted as a coherent quantum revival effect, where all the eigenmodes forming the initial state in Eq.~\eqref{eq:StateAnsatz} periodically rephase, i.e., $c_e(t= n \tau_{\rm rev})\simeq c_e(0)$ for $n=1,2 \ldots$. 

\begin{figure}
    \centering
    \includegraphics[width=\columnwidth]{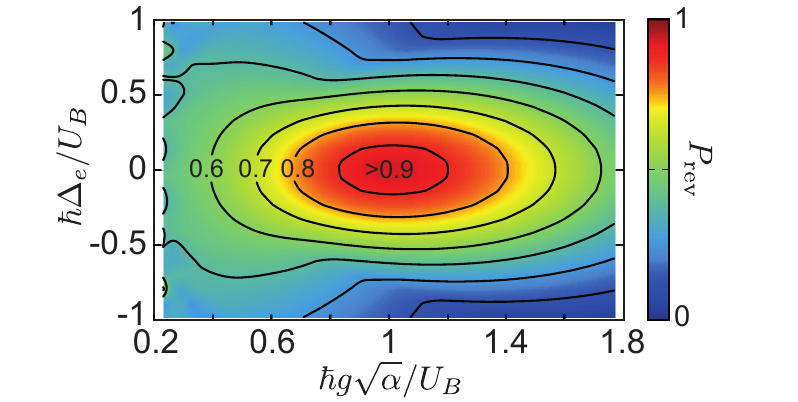}
    \caption{Maximum value of the revival probability $P_{\rm rev}$ of the excited state population after a time $t\sim \tau_{\rm rev}$. This probability is plotted as a function of the light-matter coupling $g$ and the local detuning $\Delta_e$. These results were derived  from Eq.~ \eqref{eq:eq_retarded_single_atom}, using the continuum Green's function with PBC and for $\alpha = 1/10$ and $L_y/l_0 = 200$.}
    \label{fig:5plus_plus}
\end{figure}

As mentioned above, we can understand this high reabsorption probability from the symmetry and the dispersion-free propagation of the emitted wavepacket, as shown in the right panels in Fig.~\ref{fig:5}. To see under which conditions this effect occurs, we plot in Fig.~\ref{fig:5plus_plus} the maximal revival probability $P_{\rm rev}$ as a function of the coupling strength $g$ and the (local) detuning $\Delta_e$ of the emitter from the LLL. This plot confirms that high revival probabilities of $P_{\rm rev}>0.9$ can be observed within an extended parameter regime and without the need for a precise fine-tuning of any of the system parameters.

\begin{figure}
    \centering
    \includegraphics[width=\columnwidth]{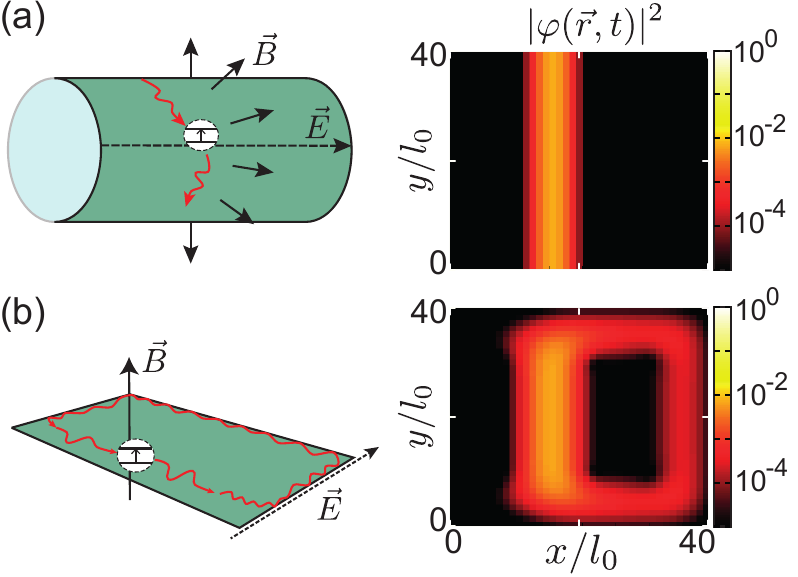}
    \caption{(a) Sketch of a lattice with PBC along the $y$-axis and OBC on the $x$-axis, and with an homogeneous out-of-plane magnetic field and an homogeneous in-plane electric field along the $x$-axis (left panel). In this case the photonic eigenmodes $f_{\lambda}(\vec{r}_i)$ are approximately described by the Landau wavefunctions in Eq.~\eqref{eq:LandauOrbitals}, which are homogeneous stripes along the $y$-direction with lateral size $\sim l_B$ (right panel). (b) Sketch of the same lattice, but with OBC along both $x$ and $y$ (left panel). In such a lattice, photonic eigenmodes form closed loops along the edge (right panel). The lattice parameters assumed for both cases are $\alpha = 1/20$ and $N_x=N_y=41$.}
    \label{fig:6}
\end{figure}

\subsection{Bulk-edge quantum channels}
\label{sec:finite_size}

Let us now switch to the experimentally more realistic scenario of a lattice with open boundary conditions. In this case the emitted photon cannot simply  return to the emitter by propagating along a straight line. Instead, once it reaches the lattice boundary, the propagation via edge-modes, which so far we have omitted from our analysis, becomes important. Surprisingly, we find that the essential features discussed for periodic systems survive for  photonic quantum Hall systems with edges. This is illustrated in Fig.~\ref{fig:6}, where we compare the characteristic shape of an eigenmode of a periodic lattice with that of a lattice with open boundary conditions. While within the bulk region, both mode functions are very similar, in the case of open boundary conditions the modefunction continues along the edges. Therefore, also in this case the photons can travel in loops and return to the emitter.   
 
\begin{figure}
    \centering
    \includegraphics[width=\columnwidth]{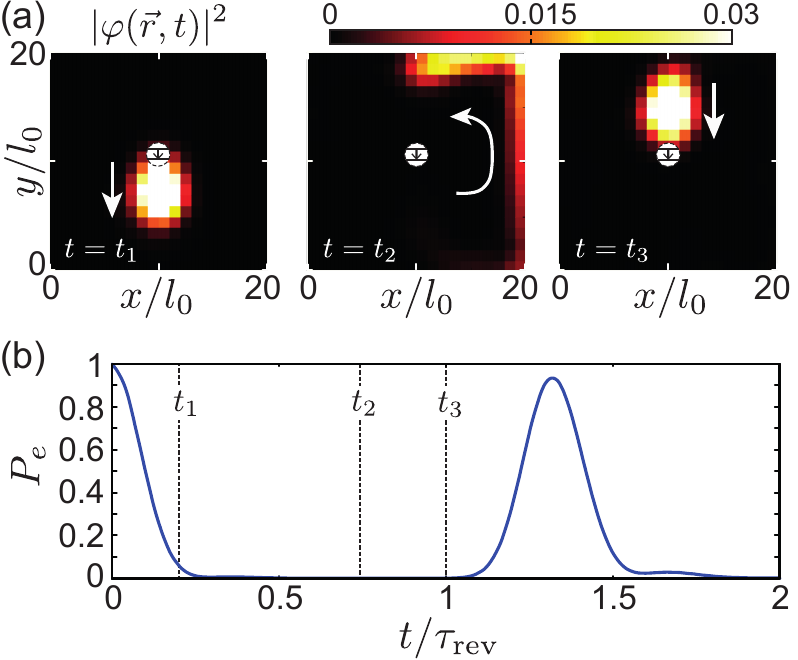}
    \caption{(a) Snapshots of the photon density $|\varphi (\vec{r},t)|^2$ at three different times after the excitation is released by an emitter located at the center of the lattice. % with initial conditions $c_e(t=0)=1$. 
    (b) Excited state population $P_e = |c_e(t)|^2$ as a function of time in units of $\tau_{\rm rev}$. The three dashed lines mark the times of the snapshots shown in (a). The other parameters for theses plots are $\alpha = 1/10$, $N_x=N_y=21$, $U_0 /(\hbar J) = 0.1$, $\Delta_e/J = 0$ and $\hbar g =  U_B/\sqrt{\alpha} \approx 0.4 \hbar J$.}
    \label{fig:7}
\end{figure}

In order to verify that this peculiar shape of the photonic eigenmodes creates an equivalence between lattices with periodic and open boundary conditions, we consider again a critically coupled emitter, but now located in the middle of a finite-size lattice. 
In Fig.~\ref{fig:7} (a) we then show three snapshot of the emitted photonic wavefunction, similar to the situation considered in Fig.~\ref{fig:5} (b) for PBC. We see that when the photon reaches the lower boundary of the system, it is transported along the edge to the upper boundary. Here, it makes a turn and propagates again through the bulk towards the emitter. 

In Fig.~\ref{fig:7} (b) we plot the corresponding time evolution of the excited state population $P_e(t) = |c_e(t)|^2$. In exactly the same way as in the case of PBC, the emitter population undergoes almost complete revivals after a time that, in spite of the much longer path, is still close to $\tau_{\rm rev}$ given in Eq.~\eqref{eq:rev_time} as a result of the much higher propagation speed along the edges.

%\subsection{Loop-states, extended states and equipotential lines}
\subsection{Guiding centers, equipotential lines and photonic demultiplexing}

The appearance of closed-loop channels, where photons propagate both in the bulk and along the edges, is puzzling at first, since it seems to break the familiar concept of topologically protected edge states.
However, the existence of such loops can be understood from an essential property of Landau wavefunctions in external potentials~\cite{Huckestein_RevModPhys.67.357}, namely that wavepackets move along \emph{guiding center} trajectories that follow the equipotential lines of this external potential landscape.

\begin{figure}
    \centering
    \includegraphics[width=\columnwidth]{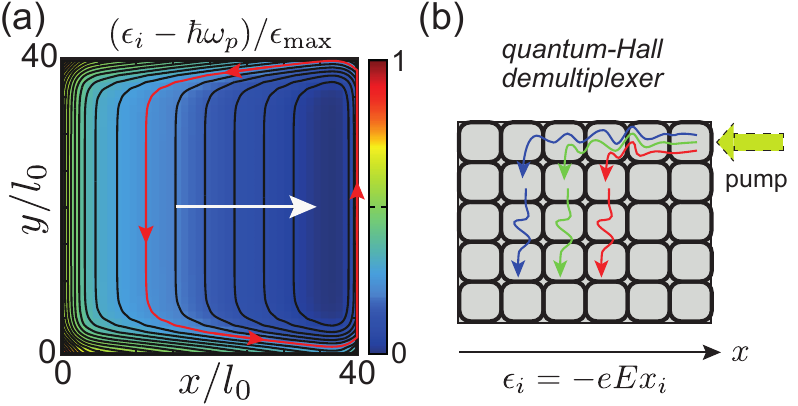}
    \caption{(a) Example of a photonic lattice with a smooth edge given by a confining potential $V_{\rm conf}$ and a linear potential gradient along the $x$-axis. The equipotential lines of the total potential (black lines) are straight lines in the bulk, but bend into a closed loop along the edge. The white arrow marks the direction of the electric field in the bulk while the red arrows show an example of a photon trajectory along the red equipotential line. (b) Sketch of the proposed photonic demultiplexing device based on the guiding-center photonic motion along equipotential lines.}
    \label{fig:8}
\end{figure}

In the bulk of the lattice, this principle readily explains the motion of the photons along straight lines, which are the equipotential lines of the linear potential gradient along the $x$-direction. The behavior of the photons near the edges can then be understood by considering the potential shown in Fig.~\ref{fig:8} (a), where on top of the constant electric field the edges are modelled by a smooth confining potential $V_{\rm conf}$, such that 
\begin{equation}
    \epsilon_i = - eE x_i + V_{\rm conf}(\vec{r}_i ).
\end{equation}
The potential $V_{\rm conf}(\vec{r}_i )$ is taken to be small and smooth in the bulk and to grow very rapidly near the boundaries of the lattice. This example illustrates very clearly how the equipotential lines in this system form closed loops, represented by straight lines in the bulk, which then continue around the edges. Therefore, these equipotential lines explain the shape of the modefunction shown in Fig.~\ref{fig:6} (b), but also the fact that the photons propagating along the edge reenter the bulk region exactly at the position $x_e$, where they have been emitted. 

In electronic systems, this guiding-center principle is crucial to understand the quantized Hall transport and the existence of extended states in the bulk, even in the presence of strong disorder~\cite{Huo_extstates_PhysRevLett.68.1375}. This latter aspect is ultimately related to the non-trivial topology of the electron wavefunctions~\cite{Arovas_PhysRevLett.60.619} and is linked to the integer quantum Hall effect through percolation theory \cite{Trugman_PhysRevB.27.7539}. Even though extended states are sensitive to the boundaries and are different for every different realization of disorder, at least one of them must always exist and connect one boundary to the other. In the current setting, this last feature is exactly what gives rise to the observed periodic photonic orbits, even in a finite-size lattice.
It is worth noticing that this type of physics was also recently observed in ultracold gases experiments with rotating traps \cite{Zwierlein_10.1126/science.aba7202, Zwierlein_nature_2022}.

As another consequence of this principle and a key difference from the $E=0$ case, photons injected at the boundary of the quantum Hall lattice can penetrate into the bulk while following specific equipotential lines. Since every equipotential line identifies a unique resonance frequency, photons of different frequency are transported to different regions in the bulk. More specifically, for a configuration shown in Fig.~\ref{fig:8} (b), photons that are injected at the upper-right corner of the lattice with a frequency $\omega_{\rm in}$ will propagate along the edge before making a turn into the bulk region at a position
\begin{equation}
    x_{\rm out} =  -l_B \, \frac{ \hbar(\omega_{\rm in} - \omega_{\rm ch}) }{U_B}\,.
\end{equation}
Therefore, this system realizes in a natural way a frequency-demultiplexing element for photons. By requiring that the separation between the output channels exceeds the spatial width of the Landau orbitals, i.e., $\Delta x_{\rm out} >l_B$, we can estimate a frequency resolution of $\delta \omega \simeq U_B/\hbar$ and a total number of $N_\omega \approx L_x/l_B$ frequency components that can be spatially separated with such a basic device.

\subsection{Quantum state transfer: edge-to-edge versus bulk-to-bulk}
\label{sec:multi_emitters_chiral_wguide}

\begin{figure}
    \centering
    \includegraphics[width=\columnwidth]{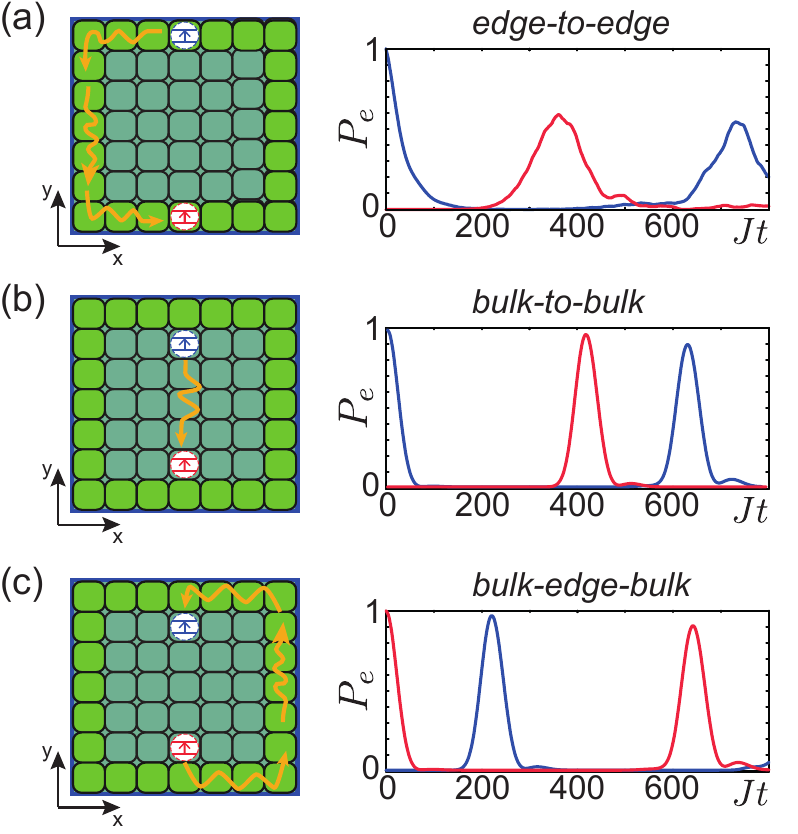}
    \caption{Chiral excitation transfer in different configurations. The left panels depict the locations of the two emitters in the lattice, while the right panels show the evolution of the excited state populations, $P_e^{(n)}(t)=|c_n(t)|^2$. (a) Excitation transfer via edge channels, where the two emitters are located on the edge (light green sites) at positions $\vec{r}_e^{\,1}/l_0 = (20,40)$ and $\vec{r}_e^{\,2}/l_0 = (20,0)$. The photon propagates along the edge from emitter $1$ to emitter $2$. The parameters in this example are $U_0/(\hbar J)= 0$, $ g_1 = g_2 = 0.1 \omega_B/\sqrt{\alpha} \approx 0.4 J $ and  $\Delta_1 = \Delta_2 = 0.1\times \omega_B \approx 0.7J$.
    (b) Excitation transfer via the bulk, where the two emitters are located at positions $\vec{r}_e^{\,1}/l_0 = (20,36)$ and $\vec{r}_e^{\, 2}/l_0 = (20,5)$. The photon propagates through the bulk from emitter $1$ to emitter $2$. 
    The parameters in this example are  $U_0/(\hbar J)= 0.05$, $\hbar g_1 = \hbar g_2 = U_B/\sqrt{\alpha} \approx 0.2 \hbar J$ and $\Delta_1 = \Delta_2 = 0$.
    (c) Excitation transfer between two emitters in the bulk, but with the photon propagating along the edge from emitter $2$ to emitter $1$.
    The other parameters are the same as in (b). 
    In all three simulations we have assumed $\alpha = 1/10$ and $N_x=N_y=41$.}
    \label{fig:10_tot}
\end{figure}

{
Let us now generalize the previous analysis to a multiple emitter case and discuss a basic application of photonic quantum Hall systems, namely to transfer an arbitrary quantum superposition state between two such emitters, i.e., to realize the mapping
\begin{equation}\label{eq:StateTransfer} 
(\alpha |g\rangle_1 +\beta |e\rangle_1)|g\rangle_2 \rightarrow |g\rangle_1(\alpha |g\rangle_2 +\beta |e\rangle_2).
\end{equation} 
Such state-transfer processes have previously been discussed in great detail for emitters coupled to chiral waveguides or to edge channels of 2D photonic lattices systems. In this case the transfer is achieved via the emission of a single photon and a high efficiency requires time-dependent couplings $g_{i=1,2}(t)$ in order to reabsorb of this photon with close to unit probability~\cite{Cirac_PhysRevLett.78.3221}. Note that state transfer schemes have also been analyzed in 1D spin chains and topological lattices  \cite{Andrew_PhysRevLett.92.187902, Suotang_PhysRevA.98.012331,longhi_https://doi.org/10.1002/qute.201800090, Chaohong_PhysRevA.101.052323, Bo_PhysRevA.106.022419}, where, however, a high-fidelity transfer again relies on very specific coupling patterns or time-dependent control techniques. Compared to those settings, the findings in the previous sections suggest that our photonic quantum Hall systems offer an essential advantage for state transfer applications, by enabling an almost perfect excitation transfer without the need for any fine tuning or additional time-dependent control.

To support this intuition, we consider a small photonic lattice as depicted in Fig.~\ref{fig:10_tot}, with two emitters located either on the edge of the lattice or in the bulk at positions $\vec r_e^{\, n}$.  By assuming an initial state as in Eq.~\eqref{eq:StateTransfer} the ansatz for the full state generalizes to 
\begin{equation}\label{eq:StateAnsatz2}
\begin{split}
|\psi\rangle(t)=& \alpha|g\rangle_1|g\rangle_2|{\rm vac}\rangle + \beta  \Big[ e^{-i\omega_{e}^1 t}c_1(t) \sigma_+^1 + e^{-i\omega_{e}^2 t}c_2(t) \sigma_+^2   \\
& + \sum_i \varphi(\vec r_i,t) \Psi^\dag(\vec r_i)\Big]|g\rangle_1|g\rangle_2|{\rm vac}\rangle,
\end{split} 
\end{equation}
with $c_1(0)=1$ and $c_2(0)=0$. Since any deterministic propagation phase can be reabsorbed into a local basis rotation, we can quantify the fidelity of the state-transfer process in terms of the excitation probability $P_e^{(2)}(t)=|c_2(t)|^2$ and define $\mathcal{F} = {\rm max}_t \,  P_e^{(2)}(t)$, given that the first emitter is initially fully excited, $P_e^{(1)}(0)=1$~\cite{Dlaska_2017,Lemonde_2019}.

After integrating out the photonic components, we obtain a coupled set of equations for the emitter amplitudes~\cite{DeBernardis_PhysRevLett.126.103603},
\begin{equation}\label{eq:eq_retarded_many_atoms}
\begin{split}
    &\dot{c}_n(t) = 
    \\
    &-\sum_{m = 1}^{N=2} \frac{g_n g_m}{4} \int_0^t ds \,G(t-s,\vec r_e^{\, n}, \vec r_e^{\, m}) c_{m}(s) e^{i(\omega_e^n t - \omega_{e}^m s)}.
\end{split}
\end{equation}
Provided the emitters are spatially separated, we can then again replace the full photonic Green's function by its continuum approximation in Eq.~\eqref{eq:Green's_fun_PBC_infinitey_equalx} and define the local detuning of each emitter as $\Delta_n = \omega_e^n - \omega_{\rm ch}(x_e^n)$. This shows that all the considerations done above for a single emitter, in particular the identification of the three different coupling regimes, remain valid for multiple emitters as well. In addition, we can numerically integrate the exact dynamics for multiple emitters within the single-excitation sector and use it to evaluate the excitation probabilities $P_e^{(n=1,2)}(t)=|c_n(t)|^2$ shown in the right panels of Fig.~\ref{fig:10_tot}.  
}

First of all, the simulations in Fig.~\ref{fig:10_tot} (a) confirm that in a conventional lattice with a synthetic magnetic field, but no synthetic electric field, $U_B=0$, excitations can be efficiently transported along the edges and even undergo a full loop without any significant losses into the bulk modes~\cite{Yao_topospinedge_2013, Dlaska_2017, Lemonde_2019, Longhi_photonics7010018, Longhi_PhysRevA.100.022123, Simon_chiralQuantumOptics_2022}. However, similar to the weak-coupling regime discussed above, the photons emitted into the edge channels have an spatially asymmetric wavefunction and, thus, are only partially reabsorbed by the second emitter, i.e., $P_e^{(2)}\lesssim  0.6$ (as also observed in \cite{Longhi_PhysRevA.100.022123}). Further, as illustrated in Fig.~\ref{fig:1} (c), the edge modes have a non-negligible dispersion~\cite{Alejandro_https://doi.org/10.48550/arxiv.2207.02090}, which is responsible for a broadening of the wavepacket during propagation and leads to a dependence of the transfer process on the distance between emitters. Therefore, the implementation of high-fidelity state transfer operations in this configuration requires additional control over the individual couplings, $g_n\rightarrow g_n(t)$, to facilitate the reabsorption process and to compensate for propagation effects~\cite{ Yao_topospinedge_2013, Dlaska_2017, Lemonde_2019}.  

In Fig.~\ref{fig:10_tot} (b) and (c) we consider an alternative setup where a sizable synthetic electric field $U_B\neq 0$ is introduced and the emitters are located in the bulk region of the lattice. In the situation assumed in Fig.~\ref{fig:10_tot} (b), where the upper emitter is initially excited, the photon propagates in a straight line through the bulk toward the second emitter. As discussed in Sec.~\ref{subsec:revivals} above, under critical-coupling conditions, the photon wavepacket in this case is highly symmetric and can be reabsorbed by the second emitter with near perfect fidelity {($\mathcal{F}\approx 0.97$)}, which is also fully reproduced by exact numerical simulations. To demonstrate a two-way connectivity, we also consider the opposite situation, where the lower emitter is initially excited. In this case, the photon must propagate along the edge, but nevertheless we observe an almost perfect transfer of the excitation. Remarkably, the broadening effect of the edge mode dispersion on the photon wavefunction is in fact of minor importance in this case of a symmetric wavefunction. 

Interestingly, in these simulations the transfer along the edge is faster than a direct transfer through the bulk. As already mentioned above, this can be attributed to the much higher group velocity in the edge channel. Indeed, the propagation time along the edges is almost negligible, compared to the propagation time in the bulk, which allows us to estimate the total transfer time by 
\begin{equation}
    \tau_{\rm T} \approx  \frac{\Delta y_{\rm PBC}}{c_H} + \frac{2}{\Gamma_e},
\end{equation}
where we have also included the emission/absorption time estimated by Eq.~\eqref{eq:decay_rate_GammE}.
Here, $\Delta y_{\rm PBC}$ is the effective distance between the emitter and the receiver under PBC, i.e., by simply ignoring the path along the edges. For example, for the two configurations considered in Fig.~\ref{fig:10_tot} (b) and (c), we would expect $J \tau_{\rm T} \approx 430$ and $J \tau_{\rm T} \approx 175$. 
In the first case the agreement with the exact simulation is almost perfect, while in the second case we have around $\sim 20 \%$ of error as the transfer time observed in the simulation is around $Jt \sim 220$.

{While an intrinsic state-transfer fidelity of $\mathcal{F}\sim 0.97$ is already impressive, it would not be sufficient for a quantum computing application and one may wonder if even higher fidelities, $\mathcal{F}\rightarrow 1$, can be achieved in principle. From the parameter scan in Fig.~\ref{fig:5plus_plus} it is clear that this is not possible for the considered purely linear electric potential. However, additional terms in the local electric field, for example of the form
\begin{equation}
    \hbar \epsilon_i = -eE x_i + \hbar \omega_p + \sum_{s}a_s \left(x_i - X_s\right)^s,
\end{equation}
do not change the overall physics of the transfer as we are going to discuss in more detail in Sec. \ref{sec:percolation_network}, but can be used to adjust the detailed shape of the local density of states and, therefore, the non-Markovian features. 

Simple numerical scans already show that by including $s=2$ terms in this expansions, fidelities  $\mathcal{F}\simeq 0.99$ can be achieved, and a further improvement is expected from a more systematic optimization. This must also include the value of the magnetic flux $\alpha$ as an optimization parameter, since the Landau levels of the Harper-Hofstadter lattice are not completely flat and, even if exponentially suppressed, residual Bloch oscillations might affect the state transfer fidelity.
This exciting perspectives go beyond the scope of the current analysis and will be the subject of future research.}

\subsection{Effect of photon losses and disorder}

\begin{figure}
    \centering
    \includegraphics[width=\columnwidth]{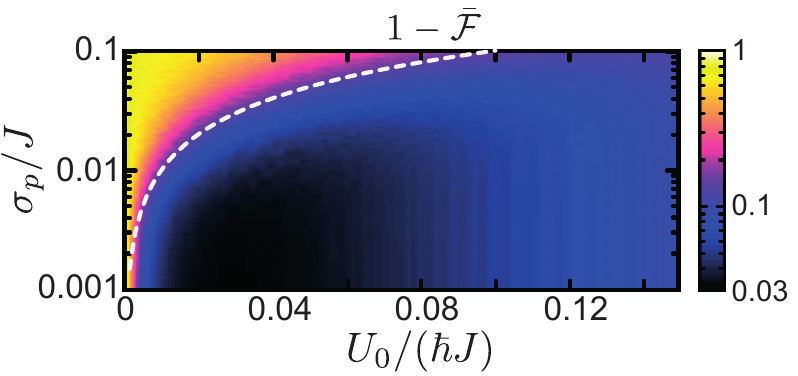}
    \caption{Disorder-averaged value of the transfer infidelity $1-\bar{\mathcal{F}}$ (color scale) as a function of the electric field, $U_0/(\hbar J)$, and the lattice disorder strength, $\sigma_p/J$. The coupling strength of the two emitters are fixed to the critical-coupling condition $\hbar g_1 = \hbar g_2  = U_B/\sqrt{\alpha}$. Emitter 1 is located at $\vec{r}_e^{\,1}/l_0 = (10, 17)$ while the emitter 2 is located at $\vec{r}_e^{\,2}/l_0 = (10, 4)$. 
    The other parameters for this simulation are $\alpha = 1/10$, $N_x=N_y=21$, $\Delta_e^1=\Delta_e^2 = 0$ and  $\gamma_p/J = 10^{-5}$ (this reduced value of the losses is chosen to highlight the effect of disorder). The dashed white line indicates the condition $\hbar \sigma_p = U_0$.
    For each choice of parameters, the disorder average is performed over $N_{\rm dis}= 100$ realizations.}
    \label{fig:11-1/2}
\end{figure}

In view of experimental demonstrations and practical applications, it is important to assess the robustness of the state transfer with respect to photon losses and a static disorder. For this purpose, we extend our study of the transfer process in Sec. \ref{sec:multi_emitters_chiral_wguide} introducing a spatial inhomogeneity of the frequency of each site of the form $\epsilon_i = -eEx_i + \hbar \omega_p^i$, where the offsets $\omega_p^i$ are chosen randomly and independently according to a Gaussian distribution with mean value $\omega_p$ and standard deviation $\sigma_p$. Then, using the same approach as in \cite{DeBernardis_PhysRevLett.126.103603}, we model the effect of photon losses with rate $\gamma_p$ by an additional damping term in the dynamics of the photon wavepacket,
\begin{equation} 
\partial_t \varphi (\vec{r}_i,t) = \ldots - \frac{\gamma_p}{2} \varphi (\vec{r}_i,t)\,.
\end{equation}
This expression already shows that the effect of photon loss only introduces an overall exponential damping of the photon amplitude, but does not affect any of the topological properties of the system~\cite{ozawa_RevModPhys.91.015006}. For state-transfer applications, it reduces the final transfer fidelity, $P_e^{(2)}(\tau_{\rm T})\approx e^{-\gamma_p \tau_{\rm T}}P_e^{(1)}(0)$, which sets a condition $\gamma_p \ll \tau_{\rm T} \lesssim \tau_{\rm rev} = c_H/L_y$ for the maximally tolerable loss rate.

%Regarding disorder, 
%we show in Fig.~\ref{fig:11-1/2} the numerical results 
%we now investigate numerically its effect on the state-transfer process between two emitters that are coupled through the bulk. 
%For this purpose, we define the fidelity of the state transfer as the maximum over time of the second emitter population $\mathcal{F} = {\rm max}_t \,  P_e^{(2)}(t)$, given that the first emitter is initially excited, $P_e^{(1)}(0)=1$.
Regarding disorder, in Fig.~\ref{fig:11-1/2} we show the disorder-averaged infidelity $1-\bar{\mathcal{F}}$ 
as a function of the inter-site voltage drop $U_0$ and the strength of the lattice disorder, $\sigma_p$.  We see that with increasing electric field, the transfer becomes increasingly robust with respect to local frequency disorder. This can be understood from the fact that for $\hbar \sigma_p \ll U_0$ the disorder is not able to  efficiently couple two neighbouring photonic eigenstates, which are spaced in energy by $\Delta E \sim U_0$. Therefore, under this condition the linear slope of the Landau levels is preserved along with all the associated transport properties. This interpretation is confirmed by the simulations in Fig.~\ref{fig:11-1/2}, where we see that the condition beyond which the quantum Hall physics is washed out by disorder is indeed given by the line $\hbar \sigma_p\simeq U_0$. 

Interestingly, the transfer fidelity $\bar{\mathcal{F}}$ starts to slowly decrease again for larger values of $U_0$. This effect, however, is not related to the disorder, but is rather caused by spurious effects due to the lattice geometry and to mixing between Landau levels whenever the electric field energy starts to be compatible with the gap between Landau levels, $U_0 \sim \hbar \omega_B$. In the considered example, we empirically found that values of $U_0 \lesssim 0.1 \hbar \omega_B$ are sufficient to suppress such imperfections.

\subsection{Experimental considerations}
Experimental realizations of topological photonic systems are currently pursued both in the optical and in the microwave domain. Focusing for concreteness on the latter case, 2D photonic lattices can be fabricated out of superconducting $LC$ resonators with frequencies of about $\omega_p/(2\pi)\approx 5-10$ GHz, tunnel couplings $J/(2\pi) \approx 100$ MHz and quality factors of $Q\approx 10^4-10^5$, which corresponds to $\gamma_p=\omega_p/Q\approx 2\pi\times 50-500$ kHz \cite{Simon_chiralQuantumOptics_2022, Painter_PhysRevX.11.011015}. By engineering a lattice with $N_x=N_y=20$ resonators along each side, a voltage drop of $U_0/J\approx 0.1$ and a magnetic flux of $\alpha=0.1$, we obtain $U_B/(2\pi\hbar) \approx 13$ MHz. Therefore, we require a coupling strength of $g/(2\pi)\approx 40$ MHz to reach the critical-coupling regime, which can be readily achieved with superconducting qubits \cite{Simon_chiralQuantumOptics_2022, Painter_PhysRevX.11.011015}. 

For this setup we then obtain a typical transport time of $\tau_{\rm T}\sim  \tau_{\rm rev}\approx 126\,J^{-1}$, such that $\gamma_p \tau_{\rm rev} \sim  10^{-2}-10^{-3}$ and the photon is able to undergo hundreds of roundtrips  before it decays. At the same time, the fabrication of superconducting resonator arrays with a frequency disorder of $\sigma_p/J \sim 10^{-2}-10^{-3}$ have already been demonstrated \cite{Painter_2018,Simon_2019_nature,Simon_PhysRevX.6.041043, Painter_PhysRevX.11.011015, Simon_chiralQuantumOptics_2022}, which means that the condition $\hbar \sigma_p< U_B$ can also be met. Therefore, we conclude that experimental realizations of such photonic quantum Hall systems are well within experimental reach.

\section{Photonic quantum Hall percolation networks}
\label{sec:percolation_network}
\begin{figure}
    \centering
    \includegraphics[width=\columnwidth]{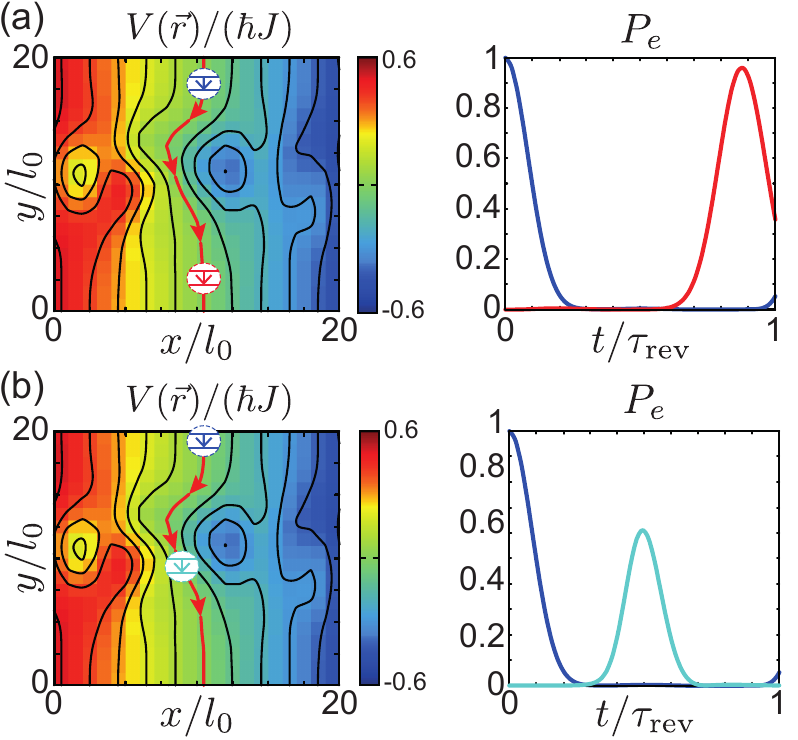}
    \caption{Chiral state transfer in the randomly generated potential $V(\vec r_i)$ shown in the two left panels. The two emitters are located along the red equipotential line, with the initially excited emitter 1 placed at $\vec{r}_e^{\, 1}/l_0 = (10,19)$. In (a) 
    the second emitter is located at a position $\vec{r}_e^{\, 2}/l_0 = (10,1)$ with the same local gradient. In this case, a perfect transfer of the excitation is possible (right panel). In (b) the second emitter is located at $\vec{r}_e^{\, 2}/l_0 = (9,9)$ in a region with a different local electric field. In this situation, the transfer probability is strongly reduced. Other parameters for these plots are $\alpha = 1/10$, $N_x = N_y = 21$, $U_0 /(\hbar J) = 0.1$, $\hbar g_1 = g_2 =  U_B/\sqrt{\alpha} \approx \, \hbar J$ and $\tilde \Delta_1 = \tilde \Delta_2 = 0$. }
    \label{fig:12}
\end{figure}

As we have already discussed above, many aspects of photon propagation in the considered quantum Hall lattice can be understood from the fact that the photonic wavepackets move along equipotential lines. As an immediate consequence, most of the effects that we have discussed so far for the case of a constant electric field, can be generalized to generic lattice potentials of the form
\begin{equation}
    \epsilon_i = V(\vec{r}_i ) + \hbar\omega_p,
\end{equation}
where $V(\vec{r}_i)$ describes a smooth, but otherwise arbitrary profile of the on-site energy offsets. In particular, the equipotential lines of $V(\vec{r}_i)$ can be curved to connect different parts of the lattice or even intersect each other. In the following we show how this additional tunability can be used to realize \emph{photonic networks}, in which many emitters can be coupled through fully configurable chiral channels. 

Interestingly, this idea is closely connected to the concept of the \emph{quantum Hall percolation network}, which was introduced for electronic systems to obtain an intuitive explanation of the integer quantum Hall effect \cite{Chalker_1988, KRAMER2005211}. In these electronic setups the network of equipotential lines is however provided by natural disorder making it impossible to control and use for any technological purpose. On the contrary, the photonic implementation allows for  almost complete freedom in the design of equipotential line, which opens up a completely new perspective for quantum Hall networks.

\subsection{Configurable chiral channels} 
In a first step, let us generalize the concept of chiral transfer channels to arbitrary potentials $V(\vec{r}_i)$, assuming, however, that variations are sufficiently smooth. In this case we can locally expand the potential around the positions $\vec r_e^{\, n}$ of the emitters,
\begin{equation}
\epsilon_i\simeq  \hbar \omega_p + V(\vec r_e^{\, n}) + \vec \nabla V (\vec{r}_e^{\, n} ) \cdot (\vec r_i-\vec r_e^{\, n}).
\end{equation} 
This simply means that each emitter sees a quantum Hall lattice with a different frequency offset and an effective field $E\sim  \vec \nabla V (\vec{r}_n )$. 
Under this assumption, the same emission and absorption dynamics discussed above is recovered if we assume a local density of states as given in Eq.~\eqref{eq:DOS_ph_approximate} and we replace the Landau voltage by
\begin{equation}
    U_B \longmapsto \tilde U_B(\vec{r}_e^{\, n}) = |\nabla V (\vec{r}_e^{\, n} )| l_B,
\end{equation}
and  the local emitter detuning by
\begin{equation}
    \Delta_n \longmapsto \tilde \Delta_n(\vec r_e^{\, n})= \omega_e^n - \tilde \omega_{\rm ch}(\vec{r}_e^{\, n})\,.
\end{equation}
Here, the channel frequency has been generalized to the $(x,y)$ space-dependent quantity according to Eq. \eqref{eq:zero-point-freq-space-dep}
\begin{equation}
  \omega_{\rm ch}(x) \longmapsto \tilde \omega_{\rm ch}(\vec{r}) =   \omega_{\ell=0}^{LL} + \frac{ V (\vec{r})}{\hbar} + \frac{\tilde U_B (\vec{r} )^2}{2\hbar^2\omega_B}.  
\end{equation}

Based on this local-field approximation, we can identify three criteria for {reaching} a state transfer between two emitters located at positions $\vec r_e^{\, 1}$ and $\vec r_e^{\, 2}$ {with similar efficiency as in Sec. \ref{sec:periodic_rev_ballistic} D}:

\begin{enumerate}
%\item The two emitters are connected by an equipotential line with energy $V(\vec r_e^{\, 1}) =V(\vec r_e^{\, 2})$.

\item The two emitters are resonant and connected by a generalised equipotential line set by the channel frequency, $\tilde \omega_{\rm ch}(\vec r_e^{\, 1}) = \tilde \omega_{\rm ch}(\vec r_e^{\, 2})$.

\item The coupling $g_n$ of each emitter satisfies the local critical coupling condition, $g_n\sqrt{\alpha}\approx\tilde U_B(\vec r_e^{\, n})/\hbar$.  

\item The local synthetic electric field is the same for both emitters, $\tilde{U}_B(\vec{r}_e^{\,1}) = \tilde{U}_B(\vec{r}_e^{\,2})$.
%The transverse width of the emitted wavepacket must match the one that a spontaneously emitted photon by the second emitter would have.

\end{enumerate}
These conditions ensure the resonant emission of a symmetric wavepacket, which can be reabsorbed by the second emitter under the same critical-coupling condition. As long as the potential does not vary too abruptly, 
the photon 
%preserves its symmetric shape during the propagation and 
moves along the equipotential line with a local Hall speed $c_H \mapsto c(\vec{r})=|\nabla V(\vec{r})|l_B^2/\hbar $ proportional to the potential gradient. 
The third criterion specifically ensures that the transverse width of the emitted wavepacket matches that of a spontaneously emitted photon from the second emitter, thus preserving the symmetry between emission and absorption.

To illustrate and validate this working principle in terms of a concrete example, we simulate a state transfer between two emitters that are coupled to a photonic lattice with the more complicated potential $V(\vec r)$ shown in Fig.~\ref{fig:12} (a). For this scenario, we compare the case where the two emitters see the same local potential gradient with the case where the gradient is different. However, in both situations the emitters are located along the same equipotential lines and are in local resonance, $\tilde \Delta_n=0$. We find that in the first case, where all three conditions from above are satisfied, the state transfer occurs with an almost perfect fidelity, despite a very complicated energy landscape. 
In the other case both emitters are critically coupled to the same equipotential line, but they are located in regions with different field gradients. Therefore, they are resonant with different channel frequencies $\tilde \omega_{\rm ch}(\vec r_e^{\, 1}) \neq \tilde \omega_{\rm ch}(\vec r_e^{\, 2})$.
In this way the absorption cannot be maximized and remains always around $\sim 60\%$ value.

\begin{figure}
    \centering
    \includegraphics[width=\columnwidth]{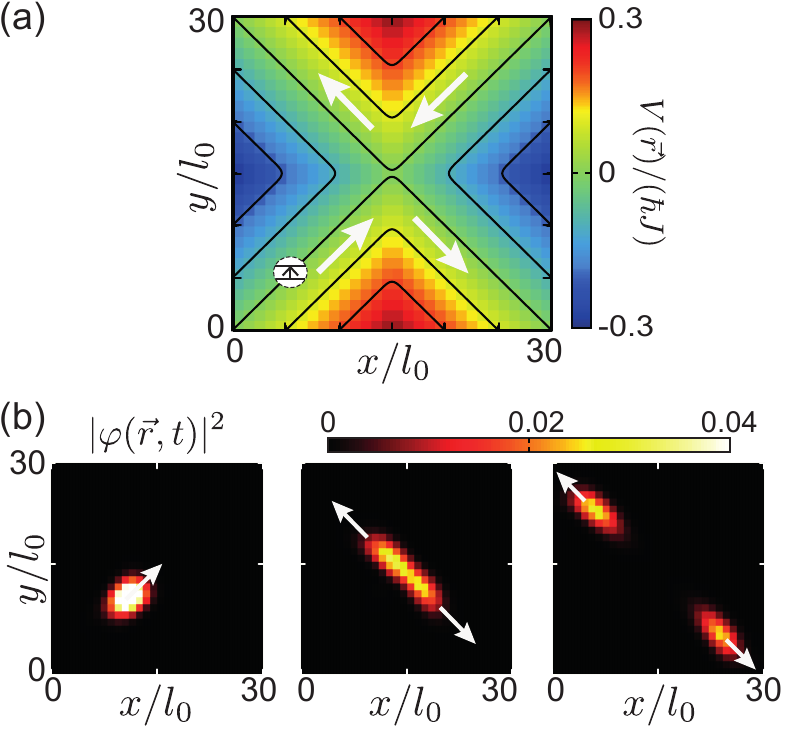}
    \caption{(a) Sketch of the lattice potential $V(x, y) = eE(\left|x-X_0\right|-\left|y-Y_0\right|)$ for the realization of a chiral beam splitter. The white arrows indicates the propagation direction along the specific equipotential line connected to an emitter located at $\vec{r}_e/l_0 = (6,6)$.
    (b) Snapshots of the photon density $|\varphi(\vec r,t)|^2$  for a photon that is emitted in the critical-coupling regime.
   The parameters for this simulation are $\alpha =1/10$, $N_x=N_y=31$, $U_0/(\hbar J) = 0.1$, $\Delta_e = 0$ and  $\hbar g = U(\vec{r}_e) /\sqrt{\alpha}$, $X_0/l_0 = Y_0/l_0=15$.}
    \label{fig:11}
\end{figure}

\subsection{Beam splitters}
In our discussion above we have implicitly assumed that the spatial variations of the applied potential are sufficiently smooth and that equipotential lines never cross. In this case the whole lattices separates into a set of disjoint 1D channels. In order to achieve higher degrees of connectivity and non-local operations, we can violate those assumptions in order to realize beam splitter elements that coherently couple different channels. 
{This is an essential element for any photonic network and is also at the core of quantum computational schemes with linear optics \cite{Milburn_RevModPhys.79.135, Mile_PhysRevA.76.032321}, provided that one can tune each network element to produce the desired output. Beam-splitter interactions also play a major role in the discussion regarding the so-called quantum Hall extended states \cite{Trugman_PhysRevB.27.7539, Arovas_PhysRevLett.60.619, Huo_extstates_PhysRevLett.68.1375} and in 
the percolation random network description of the quantum Hall effect \cite{Chalker_1988, KRAMER2005211}. Here, disorder-induced crossings of equipotential lines or the tunneling between neighbour equipotential lines give rise to effective nodes with multiple input and output channels, however, with random transmission amplitude. 
In contrast, in the current synthetic quantum Hall system, the same processes can be implemented under fully controlled conditions.} 

%provided that one can tune each network element to produce the desired output.

%it plays also a major role in the percolation random network description of the quantum Hall effect \cite{Chalker_1988, KRAMER2005211}. Here the crossing or equipotential lines or the tunneling between neighbour equipotential lines give rise to nodes with multiple input and output channels with variable transmission amplitudes. Such beam splitters are also at the core of quantum computational schemes with linear optics \cite{Milburn_RevModPhys.79.135, Mile_PhysRevA.76.032321}, provided that one can tune each network element to produce the desired output. In electronic systems this is not possible, since the network is provided naturally by disorder, but in this photonic implementation instead we have complete freedom in tunability.}

{As a proof of principle, we illustrate the the implementation of a 50/50 beam splitter that preserves the shape of the incoming wavepacket. In this way input and output photons can be critically emitted or absorbed for high-fidelty state-transfer applications. }
As illustrated in Fig.~\ref{fig:11} (a), such a beam splitter can be obtained at a crossing point of different equipotential lines, which exist, for example, for a potential that is locally of the form
\begin{equation}\label{eq:contact_point_V}
    V(\vec{r}) = eE (\left|x-X_0\right|-\left|y-Y_0\right|),
\end{equation}
{centered in $\vec{R}_0 = (X_0, Y_0)$.}
In the three plots in Fig.~\ref{fig:11} (b), we plot the propagation of a photonic wavepacket, which is emitted along the diagonal equipotential line in the north-east direction towards the crossing point at $\vec r=0$. This photon is then coherently split into two symmetric wavepackets, which propagate into opposite north-west and south-east directions. Thanks to the chirality, this process occurs without any backscattering and, furthermore, preserves the symmetric shape of the outgoing wavepackets. Therefore, the ability of being fully reabsorbed by other emitters is not degraded by this operation. This provides a great flexibility for realizing a variety of connectivity patterns in such percolation networks, in particular when beam splitters are implemented with reconfigurable potentials.   

Note that for electronic systems, a closely related splitting mechanism has been previously analyzed for quadratic saddle-point potentials, $V(\vec r)\sim x^2 - y^2$~\cite{Halperin_PhysRevB.36.7969,Champel_PhysRevB.82.161408}. However, in this case curvature effects are always relevant \cite{Champel_PhysRevB.82.045421}, making the dynamics more complicated and less controlled than for the linear saddle potential given in Eq.~\eqref{eq:contact_point_V}. 
Specifically, we have numerically observed that the quadratic saddle point does not preserve the symmetry of the photon wavepacket. Moreover, away from the contact point, our linear saddle-point potential has the same gradient in all four branches marked by the white arrows in Fig.~\ref{fig:11} (a), which ensures the conditions for high-fidelity state-transfer operations. This once more highlights the intriguing new possibilities offered by highly tunable photonic platforms, where the potential configurations can be engineered in an optimal way.

\section{Conclusions}
\label{sec:conclusions}

In summary, we introduced a new chiral quantum optics platform based on two-level emitters coupled to the bulk of a 2D photonic lattice subject to crossed synthetic magnetic and electric fields. The presence of the combined synthetic fields forces photons to propagate unidirectionally along an effective waveguide orthogonal to the electric field. 
The lateral position of the selected effective waveguide is controlled by the resonance frequency of the emitter.

Depending on the strength of the emitter-light coupling, we identified and characterized three different regimes of light-matter interactions: weak-coupling (Markovian), strong-coupling (non-Markovian) and critical-coupling (non-Markovian). The Markovian weak-coupling regime corresponds to the usual light-matter coupling regime considered in the chiral quantum optics literature, and all existing results for generic chiral setups directly extend to our system.
On the contrary, the strong-coupling and, even more, the critical-coupling regimes display radically new properties that stem from the frequency-dependent density of states: the strong-coupling regime supports atom-photon bound-states in a chiral continuum, which do not exist in standard setups.
In the critical-coupling regime, the emission process displays strongly non-Markovian features due to the interplay between light-matter interactions and quantum Hall physics.

{
The ensuing temporal symmetry of the emitted photon wavefunction can then be exploited to implement state-transfer protocols between two emitters with a fidelity that largely exceeds standard chiral quantum optics configurations, already without any fine-tuned couplings or time-dependent optimal control schemes. 
Provided that conventional pulse control strategies can still be added to correct for any of the residual absorption errors, this result can be also optimized by only adding corrections to the local synthetic electric field potential beyond the linear term.
This second way to improve the state transfer in the critical regime relies only on the reshaping of the photonic local density of states, fully exploiting the non-Markovian nature of the light-matter interactions in this quantum Hall setting. 
It thus also sets the basis to develop a new powerful mechanism to realize optimal state-transfer by engineering the local density of states of a propagating chiral channel, for which this quantum Hall setup could give the basics intuition.

}

For generic, non-uniform synthetic electric field configurations, we related the photon propagation to the fundamental property of the quantum Hall current to flow along the equipotential lines of the single particle potential according to the guiding-center motion. This can be used as a starting point for implementing new photonic devices, including frequency-(de)-multiplexing elements, chiral waveguides with arbitrary paths within the 2D lattice, and beams splitters.

{
Based on these analytic and numerical findings, we argue that all these elements can be combined to realize a full-fledged chiral quantum optical network completely based on the quantum Hall effect for light. On the one hand, such networks are of interest for technological applications, where quantum communication schemes discussed previously for 1D chiral quantum optical networks can now be extended to 2D emitter arrays and improved by the intrinsic non-Markovian dynamics of light-matter interactions in this system. On the other hand, this system is also of a purely fundamental interest, providing a new platform that extends the edge modes dynamics to the bulk and where concepts for instance related to the so-called \emph{bulk-edge correspondence} or the quantum Hall percolation theory can be further explored and expanded \cite{ozawa_RevModPhys.91.015006, KRAMER2005211}. 
Furthermore, adding non-linear quantum emitters to synthetic photonic quantum Hall systems offers completely new possibilities to access and probe the physics of  the  fractional quantum Hall effect and its chiral edge dynamics~\cite{Carusotto_NatRev_2020, Alberto_PhysRevA.107.033320}. Also in this context, the basic physical processes analyzed in this work will be relevant for developing schemes to prepare and measure strongly correlated quantum many-body states in this system.}

\acknowledgements
We are grateful to Alberto Nardin, Giuseppe Calaj\`o, and Alexander Szameit for fruitful discussions. This work was supported by the Provincia Autonoma di Trento, by the Q@TN initiative, and by the PNRR MUR project PE0000023-NQSTI. PR acknowledges support from the European Union's Horizon 2020 research and innovation programme under grant agreement No. 899354 (SuperQuLAN).

\appendix

\section{Landau photons in electric fields}
\label{app:LandauLevel_E}
The mode functions of the lattice in the continuous limit are solution of the Schr\"odinger equation \cite{DeBernardis_PhysRevLett.126.103603}
\begin{equation}
    \left[ \frac{p_x^2}{2m} + \frac{(p_y + eBx)^2}{2m} - eE x \right] \Phi_{\ell k}(\vec r) = (\omega - \omega_b ) \Phi_{\ell k}(\vec r),
\end{equation}
where $p_{x/y}=-i\hbar \partial_{x/y}$, $\omega_b = \omega_p -J/2$ and $m = 1/(2Jl_0^2)$. By making the ansatz $\Phi_{k}(\vec r) = \phi(x)\exp(iky)/\sqrt{L_y}$ 
%we have
%\begin{equation}
%    \left[ \frac{p_x^2}{2m} + \hbar \omega_B %\frac{\left( l_B k + x/l_B \right)^2}{2} - eE x \right] \phi(x) = \omega \phi(x).
%\end{equation}
%Expanding and completing the square we arrive to
and completing the square we arrive at
\begin{equation}
   \left[ \frac{p_x^2}{2m} + \frac{\hbar \omega_B}{2} \left( \frac{x}{l_B} + l_B k - \frac{U_B}{\hbar \omega_B} \right)^2 \right] \phi(x) = (\omega - \omega_H) \phi(x),
\end{equation}
where $\omega_{H} = c_H k - U_B^2/(2\hbar \omega_B)$.
The eigenfrequencies are then given by
\begin{equation}
    \omega_{\ell k} = \hbar \omega_B \left( \ell + \frac{1}{2} \right) + U_B \left( l_B k - \frac{U_B}{2\hbar \omega_B} \right),
\end{equation}
while the eigenstates are just the displaced harmonic oscillator wavefunctions
\begin{equation}
    \Phi_{\ell k}(\vec r) = \frac{\exp(iky)}{\sqrt{L_y}} \varphi_{\ell}^{\rm h.o.} \left( x + l_B^2 k - U_B/(\hbar \omega_B) l_B\right).
\end{equation}
Here
\begin{equation}
    \varphi_{\ell}^{\rm h.o.}(x) = \frac{1}{\sqrt{2^{\ell}\ell ! \sqrt{\pi}}} H_{\ell} (x/l_B) e^{- x^2/(4l_B^2)},
\end{equation}
and $H_{\ell}(x)$ is the $\ell$-th Hermite polynomial.

\bibliographystyle{mybibstyle}
 
\bibliography{references}
\end{document}